\documentclass{aa}
\usepackage{graphicx}

\newcommand{\mini}{\mbox{$M_{\rm i}$}}
\newcommand{\diff}{\mbox{${\rm d}$}}

\newcommand{\ub}{\mbox{$U\!-\!B$}}
\newcommand{\bv}{\mbox{$B\!-\!V$}}
\newcommand{\vi}{\mbox{$V\!-\!I$}}

\newcommand{\jk}{\mbox{$J\!-\!K$}}

\newcommand{\mv}{\mbox{$M_{V}$}}

\newcommand{\Ks}{\mbox{$K_{\rm s}$}}
\newcommand{\mbol}{\mbox{$M_{\rm bol}$}}

\newcommand{\dmo}{\mbox{$\mu_{0}$}}
\newcommand{\ebv}{\mbox{$E_{B\!-\!V}$}}
\newcommand{\feh}{\mbox{\rm [{\rm Fe}/{\rm H}]}}
\newcommand{\mh}{\mbox{\rm [{\rm M}/{\rm H}]}}
\newcommand{\Msun}{\mbox{$M_{\odot}$}}
\newcommand{\Teff}{\mbox{$T_{\rm eff}$}}

\newcommand{\logL}{\mbox{$\log L/L_{\odot}$}}

\newcommand{\comment}[1]{}
\newcommand{\beq}{\begin{equation}}
\newcommand{\eeq}{\end{equation}}
\newcommand{\beqa}{\begin{eqnarray}}
\newcommand{\eeqa}{\end{eqnarray}}
\newcommand{\benu}{\begin{enumerate}}
\newcommand{\eenu}{\end{enumerate}}
\newcommand{\bite}{\begin{itemize}}
\newcommand{\eite}{\end{itemize}}
        \def\smallskip{\vskip 2pt}

\begin{document}

\title{Star counts in the Galaxy}
\subtitle{Simulating from very deep to very shallow photometric
surveys with the TRILEGAL code}

\author{L. Girardi\inst{1}, 
M.A.T. Groenewegen\inst{2},
E. Hatziminaoglou\inst{3,4}, 
L. da Costa\inst{3}}
\institute{
 Osservatorio Astronomico di Trieste, INAF,
	Via Tiepolo 11, I-34131 Trieste, Italy \and
 PACS ICC-team, Instituut voor Sterrenkunde, Celestijnenlaan 200B,
        B-3001 Leuven, Belgium \and
 European Southern Observatory, Karl-Schwarzschild-Str.\
	2, D-85740 Garching bei M\"unchen, Germany  \and
 Instituto de Astrof\'\i sica de Canarias, C/ V\'\i a L\'actea s/n,
	38200 La Laguna, Spain
}

\offprints{L\'eo Girardi \\ e-mail: Lgirardi@ts.astro.it}

\date{To appear in Astronomy \& Astrophysics}

\abstract{
We describe TRILEGAL, a new populations synthesis code for 
simulating the stellar photometry of any Galaxy field. 
The code attempts to improve upon several technical aspects
of star count models, by:
dealing with very complete input libraries of evolutionary tracks;
using a stellar spectral library to simulate the photometry in
virtually any broad-band system; being very versatile allowing 
easy changes in the input libraries and in the description of
all of its ingredients -- like the star formation rate, 
age-metallicity relation, initial mass function, and geometry of
Galaxy components. In a previous paper (Groenewegen et al. 2002,
Paper~I), the code was first applied to describe the very deep 
star counts of the CDFS stellar catalogue. Here, we briefly describe 
its initial calibration using EIS-deep and DMS star 
counts, which, as we show, are adequate samples to probe both the 
halo and the disc components of largest scale heights (oldest ages). 
We then present the changes in the calibration that were necessary
to cope with some improvements in the model input data, and the
use of more extensive photometry datasets:
Now the code is shown to successfully simulate also
the relatively shallower 2MASS catalogue, which probes
mostly the disc at intermediate ages, and the immediate solar 
neighbourhood as sampled by Hipparcos -- in particular
its absolute magnitude versus colour diagram --, which contains 
a somewhat larger fraction of younger stars than deeper surveys. 
Remarkably, the same model calibration can 
reproduce well the star counts in all the above-mentioned 
data sets, that span from the very deep magnitudes of CDFS 
($16<R<23$) to the very shallow ones of Hipparcos ($V<8$). 
Significant deviations (above 50 percent in number counts) 
are found just for fields close to the Galactic Center 
(since no bulge component was included) and Plane,
and for a single set of South Galactic Pole data. 
The TRILEGAL code is ready to use for the variety of 
wide-angle surveys in the optical/infrared that will become available 
in the coming years.
}

\authorrunning{Girardi et al.}
\titlerunning{Star counts}
\maketitle

\section{Introduction}
\label{sec_intro}

\label{sec_history}

The number counts of Galactic stars in a given
bin of apparent magnitude $[m_\lambda, m_\lambda+\diff m_\lambda]$
-- where $\lambda$ stands for a passband -- and towards an element of 
galactic coordinates $(\ell,b)$ and solid angle $\diff \Omega$, is given
by the fundamental equation of stellar statistics (see Bahcall 1986
for a review)
\beq
N(m_\lambda, \ell, b) = \diff m_\lambda \int_0^\infty 
 \diff r\, r^2 \,\rho(\mathbf{r})\, 
 \phi(M_\lambda,\mathbf{r})\, \diff\Omega
\label{eq_counts}
\eeq
where $r$ is the line-of-sight distance, and $\rho(\mathbf{r})$ 
is the stellar density as a function of the position 
$\mathbf{r}=(\ell,b,r)$. 
$r$, when measured in parsecs, is related to the absolute and 
apparent magnitudes $M_{0,\lambda}$ 
and $m_\lambda$, and to the interstellar absorption $A_\lambda$, by
\[ 
M_{0,\lambda}=m_\lambda-5\log r-A_\lambda(r)+5 \,.
\]
$\phi(M_{0,\lambda},\mathbf{r})$ is the intrinsic distribution of
stellar absolute magnitudes, i.e. the intrinsic luminosity function 
(LF) of the stars considered at $\mathbf{r}$.

To describe the stellar densities $\rho(\mathbf{r})$ for the 
largest possible volume, and to a lesser extent 
also $\phi(M_\lambda,\mathbf{r})$, is the ultimate task of the
so-called Galaxy star count models. To achieve these goals, 
the usual way is
to assume the functional forms of $\rho$ and $\phi$, and then
compare the results of Eq.~(\ref{eq_counts}) to observed
number counts in several Galaxy fields. A number of assumptions help
in simplifying the task. The first one is to
recognize that the Galaxy can be separated in a few distinct components,
such as the disc, halo, and bulge:
\beq
\rho = \rho_{\rm d} + \rho_{\rm h} + \rho_{\rm b} \, ,
\eeq
each one of these components having a simple expression for their 
density. The second one is to assume an intrinsic LF $\phi$ which is 
virtually independent of $\mathbf{r}$, i.e. 
$\phi(M_\lambda,\mathbf{r})=\phi(M_\lambda)$, for each component.

The recipes for $\phi(M)$ can be of two types. Either one 
(1) assumes an empirical $\phi(M)$, derived from e.g. star counts 
in globular clusters or in the Solar Neighbourhood, or 
(2) assumes a theoretical $\phi(M)$, derived from a set of 
evolutionary tracks together with suitable distributions 
of stellar masses, ages, and metallicities. 

Option (1) was the preferred one in the past, 
and the one adopted in some of the most successful Galaxy models.
Despite their success in reproducing several sorts of data, 
it is not difficult to find points of inconsistency 
in many of such models. 
A common approximation, for instance, has been that disc 
stars of different absolute magnitude present different 
scale heights (and hence $\rho_{\rm d}(\mathbf{r})$): 
Bahcall \& Soneira (1980, 1984; and many authors later on) 
assigned the scale height of 325~pc for $M_V>5.1$, 
of 90~pc for $M_V<2.3$ dwarfs, 
and linearly interpolated in between. This separation 
was interpreted as a coarse separation into 
``young'' and ``old'' populations. Red giants instead 
were assigned 250 pc. Gilmore \& Reid (1983)
adopt similar approximations for main sequence stars. 
M\'endez \& van Altena (1996, 1998) do the same, and
moreover assume a unique scale height for all evolved stars 
(subgiants, giants, and white dwarfs).
From the point of view of stellar and 
population synthesis theories, these approximations are
clearly not justified, for a series of reasons:
(a) Most coeval stellar populations 
contain both red and blue stars characterized by 
initial mass values which are, at least for the most luminous objects, 
very much the same; it is then very unlikely that the spatial 
distribution of these red and blue stars
could be different. (b) Similarly, young stellar populations
contain both bright and faint main sequence stars, whose relative
scale heights cannot change that dramatically in time-scales of less
than one Gyr. (c) Moreover, it is remarkable that population synthesis 
theory indicate that in star-forming galaxy components (e.g. in
the thin disc), red giants are 
relatively young -- most having less than say 2 Gyr
(see Girardi \& Salaris 2001) -- and not old objects;
applying the largest scale heights to all giants 
is then simply wrong.

Although this sort of inconsistency is not inherent to method (1),
they are completely removed by the use of method (2). 
In the latter, at any $\mathbf{r}$ the relative numbers of stars with 
different colours and absolute magnitude strictly obey the constraints 
settled by stellar evolution and population synthesis theories;
hence, $\rho(\mathbf{r})$ cannot be arbitrarily changed 
as a function of absolute magnitude. On the other hand, 
method (2) allows $\rho(\mathbf{r})$ to be easily expressed 
as a function of other stellar parameters, such as  
age and metallicity -- something not possible with method (1) 
where individual stellar ages and metallicities 
are not available -- then allowing the simulation of important
effects like metallicity gradients, scale lengths increasing 
with age, etc. This turns out to be a significant advantage 
of method (2) over (1).

Models that follow method (2) 
may be put under the generic name of ``population synthesis 
Galaxy star count models'', and have been developed in the
last decades by e.g. Robin \& Cr\'ez\'e (1986), Haywood (1994),
Ng et al. (1995), Castellani et al. (2002), and Robin et al. (2003). 
These works benefit from the releases of extended databases of stellar 
evolutionary tracks to predict the properties of
stars of given mass, age, and metallicity. Some assumptions then are 
necessary to give the distributions of these stellar parameters.
Such distributions may be derived, for instance, starting
from an initial mass function (IMF), an age-metallicity relation (AMR),
and a law for the star formation rate (SFR) as a function of Galaxy age.

In the present paper, we will describe a Galaxy model developed
according to the population synthesis approach, taking particular care
in the consistency among the different sources of input data. 
It has been 
developed with a primary task in head, which is, essentially:
to be capable of simulating the same sort of data that will be
released by some major campaigns of wide-field photometry 
conducted these years. Of primary importance in this context, 
are the several parts of
the ESO Imaging Survey (EIS; Renzini \& da Costa 1997), the
Two Micron All Sky Survey (2MASS; Cutri et al. 2003), 
and the Sloan Digital Sky Survey (SDSS; York et al. 2000).
Present and future data from HST deep fields,
VIMOS, VISTA, UKIDSS, GAIA, might be considered as well. 
Moreover, our model should also be able to take advantage of
the extraordinary constraints provided by the astrometric
mission Hipparcos (Perryman et al. 1997).
Of course, a program which meets these aims can be applied to
any other sort of wide field data as well. 


Before proceeding, let us briefly summarize our primary 
objectives and how these translate into technical requirements.

First of all, our primary goal is to simulate the expected 
star counts in several
passband systems, such as those used by Hipparcos, EIS, 
2MASS, SDSS, etc. For doing so, we should be able
to consistently predict the stellar photometry in a lot of 
different photometric systems. The way out to  
this problem has been settled in a previous work (Girardi et
al. 2002), that describes a quite general method for
performing synthetic photometry and deriving bolometric 
corrections from an extended library of stellar spectra.
Such tables are now routinely produced for any new system
that we want to compare the models with.

The second requirement is of being able to simulate both 
very shallow -- but of excellent quality --
photometric data samples as Hipparcos (Perryman et al. 1997), 
and very deep ones such as the EIS Deep Public Survey 
(e.g. Paper~I).
The former case implies that we include all important 
evolutionary sequences, such as most of the main sequence, 
and giants both in the red giant branch (RGB) and red clump,
which make the bulk of the Hipparcos colour-magnitude diagram (CMD).
In the latter case, we should also include an extended 
lower main sequence, reaching down to visual absolute
magnitudes as faint as $M_V\sim30$, which corresponds to 
stellar masses of $\sim0.1$~\Msun. 
Moreover, old white dwarfs start to
become frequent at such faint magnitudes as well. 
Therefore, the libraries of stellar data should be extended 
to the intervals of very low masses, and to very old white 
dwarfs.

It is also clear that these requirements imply, necessarily,
that we opt for the population synthesis approach. In fact,
there is little hope that we could collect empirical
data for such a variety of stars, in the several photometric 
systems involved, and with good enough statistics that the 
intrinsic CMDs could be constructed with reliability. 
Theoretical data, instead, is available for all of 
our purposes, as will be shown in what follows.
Such theoretical data is also routinely submitted to stringent 
tests against photometric data, such as the Hipparcos CMD,
star clusters, eclipsing binaries, red giants 
with measured diameters, etc.
In general, the errors detected in the models amount to less
than a few tenths of a magnitude, and just for some particular
stars and/or passbands. Certainly, the time is ripe for 
completely relying in theoretical data in Galaxy star count
models.

	\begin{figure*}
	\begin{minipage}{0.65\textwidth}	
	\resizebox{\hsize}{!}{\includegraphics{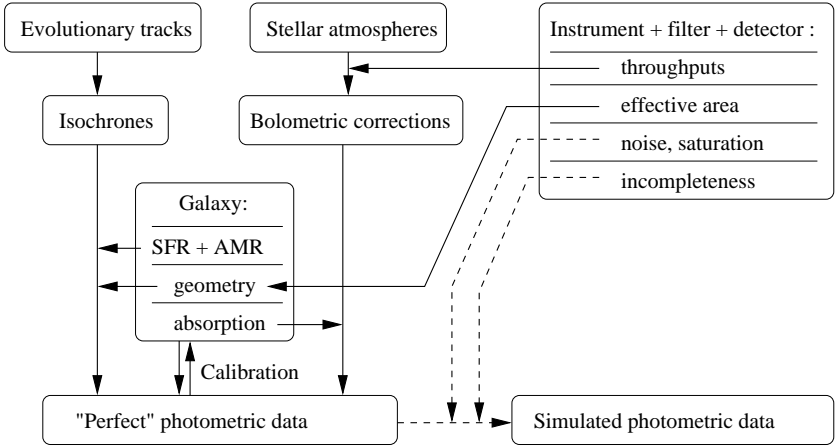}}
	\end{minipage}	
	\hfill
	\begin{minipage}{0.32\textwidth}	
	\caption{A general scheme of our codes. The continuous arrows
refer to steps which are performed inside the TRILEGAL main code 
and subroutines; they lead to the simulation of 
perfect (i.e. without errors) photometric data. The dashed lines 
refer to some optional steps, usually performed with
external scripts, like e.g.\ those mentioned in Paper~I, 
to generate catalogues with errors.}
	\label{fig_scheme}
	\end{minipage}	
	\end{figure*}

The plan of this paper is as follows:
Section~\ref{sec_code} and \ref{sec_inputdata} detail the code 
and its input data, respectively.
Sect.~\ref{sec_calib} describes its initial calibration
as performed for Groenewegen et al.'s (2002) work,
based mostly on EIS and DMS data.
Sects.~\ref{sec_2mass} and \ref{sec_hipparcos}
details and presents the fine-tuning of the initial calibration, 
including additional comparisons with 2MASS and Hipparcos data.
Sect.~\ref{sec_conclu} draws a few final comments
and summarizes the main results of the present paper.

\section{The code}
\label{sec_code}

We describe here all the necessary input for
computing a Galactic Model. Normally, this means
simulating the photometric properties of stars located towards
a given direction $(\ell,b)$. This task is performed by the newly
developed code TRILEGAL, which stands for TRIdimensional
modeL of thE GALaxy\footnote{TRILEGAL is also a word commonly
used to say ``very nice'' in southern Brazil.}.

The code is written in C language. Its core is made of a few
subroutines that efficiently interpolate and search for stars of
given mass, age, or metallicity, inside a database of stellar
evolutionary tracks. They deal with 
all the intrinsic properties of stars -- luminosity, 
effective temperature, mass, metallicity, etc. 
These subroutines, developed in
Girardi (1997), have so far been used in a series of works 
dealing from the construction of theoretical 
isochrones (e.g. Girardi et al. 2000; Salasnich et al. 2000) 
to the simulations of synthetic 
CMDs for nearby galaxies (e.g. Girardi et al. 1998; Girardi 1999, 
Girardi \& Salaris 2001; Marigo et al. 2003). 
Another set of routines, more recently developed, 
deal with all aspects related 
with synthetic photometry, i.e. the conversion between intrinsic
stellar properties and observable magnitudes. They rely on the same 
simple formalism described in Girardi et al. (2002).
  
\subsection{Scheme}
\label{sec_scheme}

A general scheme of the code is provided in
Fig.\ref{fig_scheme}. It makes use of 4 main elements: a library
of theoretical evolutionary tracks, a library of synthetic spectra, some
parameters of the detection system, and the detailed description of
the Galaxy components. The libraries of evolutionary 
tracks and spectra can be pre-processed in the form of theoretical
isochrones and tables of bolometric corrections, so as to
reduce the number of redundant operations during a
simulation. These are to be considered as ``fixed input'', but
can be easily changed so as to consider alternative sets of
data.

The instrumental setup specifies, among others:
(1) The set of filters+detector+telescope throughputs in which the
observations are performed; any change of them implies the
recalculation of the bolometric correction tables;
(2) The effective sky area to be simulated; the number of simulated 
stars scales with this quantity.

The several Galactic components (halo, thin and thick disc, bulge, 
etc) are specified by a their initial distributions of
stellar ages and metallicities (SFR and AMR), masses (IMF), 
space densities, and interstellar absorption. This is done
separately for each component. 
The space densities are in the form of simple expressions containing 
a few modifiable parameters, to be specified in 
Sect.~\ref{sec_inputdata} below.

\subsection{The simulation and output}

A run of TRILEGAL is formally a Monte Carlo simulation in which
stars are generated according to the
probability distributions already described. 
Eq.~(\ref{eq_counts}) is used for predicting the number of
expected stars in each bin of distance modulus. For each simulated
star, the SFR, AMR and IMF are used to single out the stellar age, 
metallicity, and mass. Finally its absolute photometry is derived 
via interpolation in the grids of evolutionary tracks (or isochrones),
and converted to the apparent magnitudes using the suitable values of 
bolometric corrections, distance modulus and extinction.

During the simulations, a lot of different stellar parameters
can be kept in memory and printed out: initial and current
mass, age, metallicity, surface chemical composition and gravity,
luminosity, effective temperature, core mass, etc. In the case of 
thermally-pulsing (TP-) AGB stars, this information is also used to 
simulate the pulse cycle variations (see Marigo et al. 2003).

The calculation initially produces a ``perfect
photometric catalogue'', which perfectly reflect the input 
probability distributions but for the Poisson noise. 
This catalogue can be later degraded by using the known 
photometric errors, such as photon noise, saturation and 
incompleteness for a given instrumental setup. This second task 
does not belong to TRILEGAL, but is performed by separated
subroutines, like for instance the ones described by 
Paper~I.

\section{The input datasets}
\label{sec_inputdata}

There are essentially 5 different input datasets in TRILEGAL:
      \benu
      \item tables of stellar evolutionary tracks, that give their basic 
properties (bolometric magnitude \mbol, effective temperature \Teff, 
surface gravity $g$, core mass, surface chemical composition, etc) 
as a function of initial mass \mini, stellar 
age $\tau$, and metallicity $Z$.
      \item tables of bolometric corrections $BC_{\lambda_i}$ for the 
several filters pass-bands $\lambda_i$, as a function of \Teff, 
$\log g$, and \mh, as well as the relative absorption in the several 
passbands with respect to $V$, $A_{\lambda_i}/A_V$;
    \item the IMF $\phi_m$;
    \item the star formation rate as a function of age, $\psi(t)$,
and age-metallicity relation, $Z(t)$, for the different Galaxy 
components;
	  \item the geometry of the Galaxy components, i.e. the stellar 
density $\rho(\mathbf{r})$ and differential $V$-band extinction 
$\diff A_V(\mathbf{r})$ as a function of position $\mathbf{r}$.
	   \eenu
They will be discussed separately in the following.

\subsection{Evolutionary tracks }
\label{sec_tracks}

	\begin{figure*}
	\begin{minipage}{0.65\textwidth}	
	\resizebox{\hsize}{!}{\includegraphics{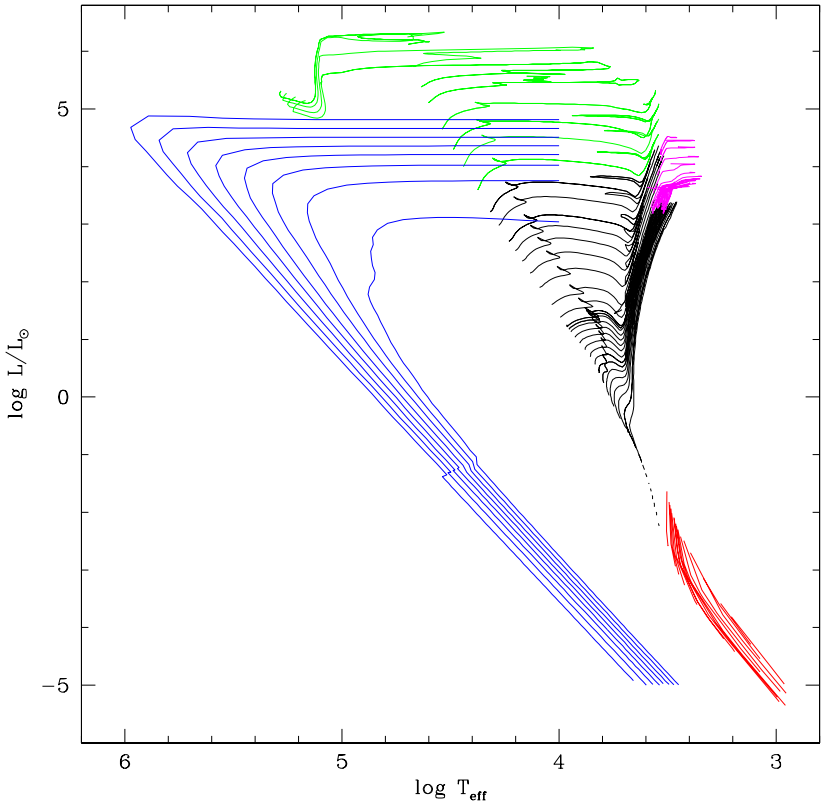}}
	\end{minipage}	
	\hfill
	\begin{minipage}{0.32\textwidth}	
	\caption{HR diagram containing all tracks assembled for
the solar metallicity. Our database contains similar data for 5 other
values of metallicity. In the electronic version of this paper,
tracks from different sources are marked with different colours: 
Girardi et al. (2000; black) for most evolutionary phases of low- and 
intermediate-mass stars, complemented with the TP-AGB phase from 
Marigo et al. (2003, and in preparation; magenta), massive stars from 
Bertelli et al. (1994; green), very-low mass stars and brown dwarfs
from Chabrier et al. (2000; red), post-AGB and PNe nuclei from
Vassiliadis \& Wood (1994) complemented with WD cooling sequences
from Benvenuto \& Althaus (1999; both in blue).
}
	\label{fig_hrd}
	\end{minipage}	
	\end{figure*}

Based on our previous work on simulations of synthetic CMDs for 
Local Group galaxies, we have assembled a large, quite complete, 
and as far as possible homogeneous -- in terms of their input 
physics -- database of stellar tracks. They are illustrated in the
HR diagram of Fig.~\ref{fig_hrd}:

1) For masses between 0.2 \Msun\ and 7 \Msun, we use the tracks from
Girardi et al. (2000), that range from the zero-age main sequence (ZAMS)
up to either the end of the TP-AGB, or to an age of 25 Gyr for the 
lowest-mass stars. Metallicities are comprised between 
$Z=0.0004$ ($\mh=-1.7$) and $0.03$ ($\mh=+0.2$). Another set with
$Z=0.0001$ ($\mh=-2.3$) and computed with the same input physics
(Girardi, unpublished) is included.

The TP-AGB evolution included in these tracks is estimated
from a simplified synthetic evolutionary code (cf. Girardi \& Bertelli
1998, case of eq. 17 plus eq. 20). Although this TP-AGB evolution
is very approximated, it provides a reasonable
initial-final mass relation (see figure 2 in Girardi \& Bertelli 1998),
and hence reasonable masses for the white dwarfs to be considered 
below. The maximum mass of WDs attained is about 1.2 \Msun\ for the 
lowest metallicities, and 0.9 for the highest.
On 2002, we replaced these simplified TP-AGB tracks for more 
detailed ones computed by Marigo et al. (2003; and in preparation), 
which are based on the new formulation for molecular opacities by 
Marigo (2002).

2) For masses lower than 0.2 \Msun, and down to 0.01~\Msun, we
include the models for very-low mass stars and brown dwarfs with
dusty atmospheres from Chabrier et al. (2000).
This provides us with a main sequence (MS) going down to
luminosities as faint as $\log(L/L_\odot)=-5$,
and to effective temperatures as cold as 916~K.
For $M<0.2$ \Msun, Chabrier et al. (2000) tracks exist only
for solar metallicity, which are hence adopted for all
metallicities in our models. It is also worth remarking that
Chabrier et al. tracks evolve on time-scales of some Gyr,
and include a significant fraction of the pre-MS evolution.

3) Post-AGB and white dwarfs tracks
of a suitable mass are attached at the end of the 
TP-AGB phase of all stars of initial mass between 0.6 and 5 \Msun.
We use the post-AGB and PNe nuclei tracks from 
Vassiliadis \& Wood (1994) down to $\logL\simeq-1.5$. 
We then shift to the WD cooling
tracks of Benvenuto \& Althaus (1999)
with CO cores, total mass ranging from 0.5 to 1.1 \Msun, a
$10^{-6}$ \Msun\ envelope layer of hydrogen, and non-zero metallicity,
plus their 1.2 \Msun\ model of zero-metallicity. 
Unfortunately, these tracks end at $\log(L/L_\odot)=-5$, 
which implies maximum WD ages of about 10 Gyr. To overcome this 
problem, we have artificially extended the tracks up to ages of
15~Gyr\comment{ by means of simple extrapolation}. 
Notice that no track is available for WDs between 1.2 and 
1.4~\Msun\ (the Chandrashekhar mass). This is not a problem at all, 
since from our AGB tracks we predict no WDs within this range 
of masses.

4) For masses higher than 7 \Msun\ we adopt the same tracks 
as in Bertelli et al. (1994) isochrones, but for $Z=0.0001$ 
and $Z=0.001$ where more recent models (Girardi et al. 1996, 
and 2003, respectively) are used.

We interpolate among all these tracks in order to derive
stars with intermediate values of mass, age, and metallicity. 
All interpolations
are performed between points of equivalent evolutionary status,
as usual in the codes for generating isochrones (e.g. Bertelli et al.
1994; Girardi et al. 2000). Linear 
interpolations are adopted, with $\log m$, $\log t$, and \mh\
being the independent variables. 

The complete set of stellar models for solar metallicity is plotted
in the HR diagram of Fig. \ref{fig_hrd}. Similar 
grids of tracks are available also for metallicities
$Z=0.0001$, 0.0004, 0.001, 0.004, 0.008, and 0.03 (limited to 
$M\le7$~\Msun\ in the latter case). 

Finally, we remark that the present stellar database corresponds
to the ``basic set'' of isochrones as mentioned by 
Girardi et al. (2002) and available at the web page
\verb$http://pleiadi.pd.astro.it/isoc_photsys.00.html$, 
but for three important improvements: the inclusion of post-AGB
and white dwarf 
cooling tracks, the extension of very-low mass stars and 
brown dwarfs down to 0.01 \Msun, and the improved 
prescriptions for the TP-AGB phase.

\subsection{Tables of bolometric corrections and absorption coefficients }
\label{sec_bc}

Once a star of $(L,\Teff,\mh)$ is selected by the code, its
bolometric luminosity $\mbol=-2.5\log L + \mbol_\odot$ is converted
into absolute magnitudes by means of $M_\lambda = \mbol - BC_\lambda$.
The bolometric corrections $BC_\lambda$ are derived from a large 
database of synthetic and empirical spectra, according to the synthetic
photometry procedure throughly descrived in Girardi et al. (2002).
Importantly, this allows the application to a very wide set of
photometric systems, provided that we deal with
\bite
\item intermediate- to broad-band filter sets (otherwise
errors in synthetic colours become significant);
\item VEGA, AB, or ST magnitudes, or systems in which 
some of the photometric standard stars are also
well-measured spectrophotometric standards.
\eite

Regarding the spectral library in use, it is essentially the same one 
described in Girardi et al. (2002), but now complemented with white
dwarf spectra. Namely, the sources of spectra are:
\benu
\item For most stars, the ``non-overshooting'' set of
models computed by Castelli et al. (1997) with the Kurucz (1993) 
ATLAS9 code (see also Bessell et al. 1998). 
This set covers \Teff\ from 50\,000 down to about 3\,500 K 
(i.e. from O to early-M spectral types), and metallicities \mh\ from 
$+0.5$ to $-2.5$. 
\item Blackbody spectra for stars with $\Teff>50\,000$ K.
\item Empirical spectra for M giants (Fluks et al. 1994).
\item ``BDdusty1999'' spectra for all dwarfs cooler than 3\,900 K, 
and down to 500 K (Allard et al. 2000).
\item
Synthetic DA white dwarf atmospheres from Finley et al. (1997)
and Homeier et al. (1998), for \Teff\ between 100\,000 and 5\,000 K.
\eenu

In addition to the BC values for stars of any $(\Teff, \log g, \mh)$,
we also compute, for any photometric system, the ratio between the
absorption coefficient in each filter, and the total absorption in
the $V$ band, $A_\lambda/A_V$. As it is well known (see e.g.\ Grebel \&
Roberts 1995; Girardi et al. 2004), this ratio is 
not strictly constant but depends on the 
spectral type of the star and the extinction curve under consideration.

In the present work we compute the $A_\lambda/A_V$ ratio for a 
G2V star (the Sun) subject to mild absorption ($A_V<0.5$ mag) and
following the Cardelli et al. (1989) absorption curve with $R_V=3.1$.
The derived $A_\lambda/A_V$ quantities\footnote{
Extinction coefficients for the Johnson-Cousins-Glass and 
SDSS photometric systems are tabulated in Girardi et al. (2004).}
are then applied to stars
of all spectral types and reddening values, although, formally,
they are adequate only for low-reddening G2V stars in the case 
$R_V=3.1$. This approach is adopted just for the sake of simplicity.
In alternative, it is very easy to implement a more accurate
approach to the problem, which will be 
followed in future applications.

\subsection{The initial mass function}
\label{sec_imf}

The IMF $\phi_m$ is a crucial ingredient because it determines the 
relative numbers of very-low mass stars, that may dominate star 
counts at visual magnitudes fainter than $\sim22$.
We have introduced the IMF in a very flexible way, so that 
it that can be easily changed. 
In order to be able to use star formation rates in units of \Msun/yr,
our default IMF normalization is for a total mass equal to 1, i.e.
\beq
\int_0^\infty m\phi_m\diff m = 1 \, \Msun\,\,\,\,.
\label{eq_imfnorm}
\eeq

Our default IMF is a Chabrier (2001) log-normal function,
\[
\phi_m \propto m\,\exp
	\left[-\frac{(\log m - \log m_0)^2}{2\sigma^2}\right]\,\,\,,
\]
whose parameters are a characteristic mass, $m_0=0.1$~\Msun, and 
dispersion, $\sigma=0.627$.

Other commonly-used IMFs, like segmented power-laws (Salpeter 1955;
Kroupa 2001) and Larson's (1986) exponential form, are also 
included in the code.

The IMF as given above refers to the mass distribution of
single stars. Additionally to them, it is very easy to 
simulate non-interacting binaries in our simulations.
When so required, we adopt the same prescription as in
Barmina et al. (2002): for each primary star of
mass $m_1$, there is a probability $f_{\rm b}$ that it 
contains a secondary, whose mass
$m_2$ is given by a flat distribution of mass ratios comprised 
in the interval $[b_{\rm b},1]$. This prescription is
particularly useful for simulating the binary sequences which
are often evident in CMDs of open clusters. Typical 
values of $f_{\rm b}$ and $b_{\rm b}$ are 0.3 and 0.7, 
respectively, which we adopt as a default.

\subsection{Star formation rates and age-metallicity relations}
\label{sec_sframr}

Each Galaxy component is made of combination of
stellar populations of varying age and metallicity.
In our code, their distribution is completely specified by the
functions SFR, $\psi(t)$, and AMR, $Z(t)$. 
Both are given in a single input file containing, for each 
age value,
\bite
\item the SFR $\psi(t)$ (in units proportional to \Msun/yr),
\item the mass fraction of metals $Z$,
\item the logarithmic dispersion of $Z$, $\sigma(\log Z)$.
\eite
Whenever necessary, $Z$ is converted into the logarithmic 
metal and iron contents by means of the approximate relations
\beqa
\mh & = & \log(Z/Z_\odot) \,, 
\label{eq_mh}\\
\feh & = & \log(Z/Z_\odot)+[\alpha/{\rm Fe}] \,,
\label{eq_feh}
\eeqa
where $[\alpha/{\rm Fe}]$ is the degree of enhancement of
$\alpha$ elements with respect to scaled-solar compositions,
and $Z_\odot=0.019$ as in Girardi et al.~(2000).
These relations provide \mh\ and \feh\ values accurate to within 
$\sim0.03$ dex, which is good enough for our purposes.

For this paper, $\alpha$ enhancement is taken into consideration
for the low-metallicity Galaxy components (halo and old disc). 
In these cases, we can safely 
convert a given \feh\ into $Z$ by means of Eq.~(\ref{eq_feh}), 
and associate to that $Z$ the evolutionary tracks computed
with scaled-solar compositions using the relations provided
by Salaris et al.~(1993). For
metallicities higher than about half solar, this approximation
is no longer valid and it is preferable to use tracks 
specifically computed for $\alpha$-enhanced composition (see
Salaris \& Weiss 1998; VandenBerg et al.~2000; Salasnich et 
al.~2000).

	\begin{figure*}
	\begin{minipage}{0.48\textwidth}	
	\resizebox{\hsize}{!}{\includegraphics{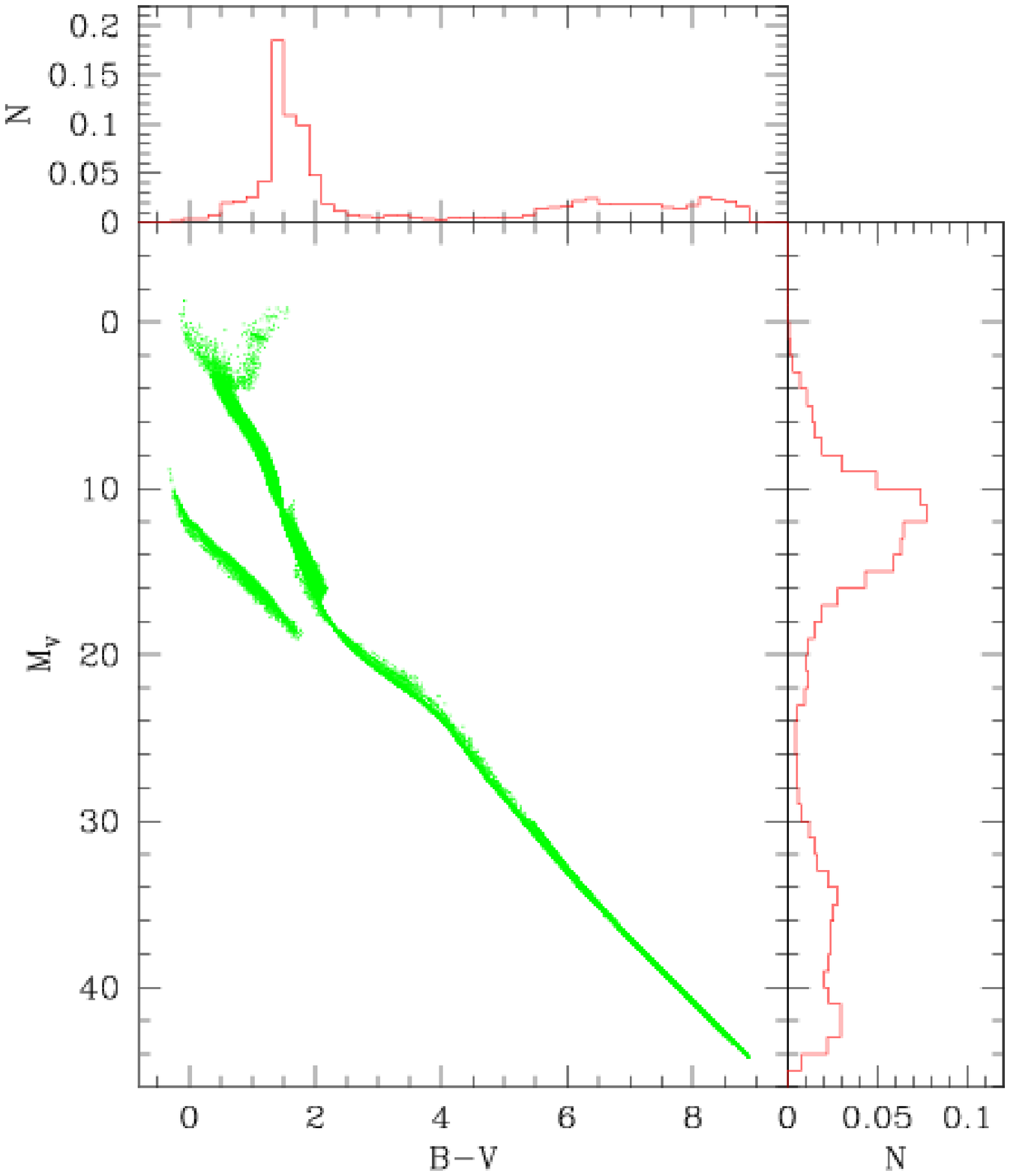}}
	\end{minipage}	
	\hfill
	\begin{minipage}{0.48\textwidth}	
	\resizebox{\hsize}{!}{\includegraphics{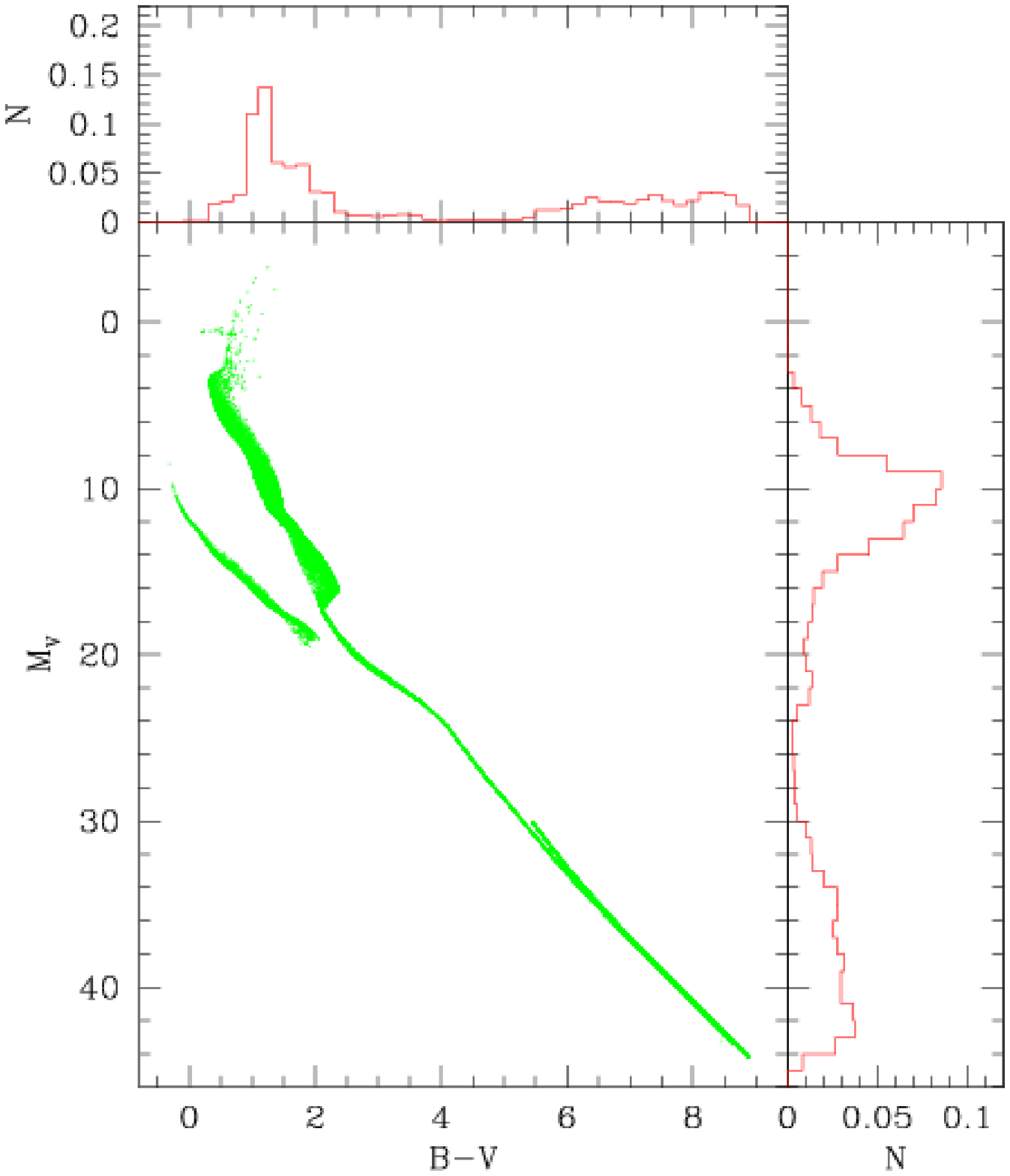}}
	\end{minipage}	
	\caption{Intrinsic $M_V$ vs. \bv\ diagrams that follow from 
our default choices of SFR, AMR, IMF, evolutionary tracks and spectra,
shown for both the disc (left) and halo (right). In both cases, the
histograms to the right and top show the corresponding luminosity and 
colour functions, respectively. Each diagram contains about $10^5$ 
simulated objects.}
	\label{fig_cmd}
	\end{figure*}

The SFR can only be considered well-known for the old Galactic 
components. In fact, for ages close to 12 Gyr, a change of age of 
1 Gyr causes just small changes in the 
stellar luminosity function, and hence has a minor impact on
the simulated star counts.
For the disc components, the SFR is less constrained.
Anyway, the Galactic model is also relatively insensitive
to the disc SFR, at least as long as we are not sampling 
regions at low Galactic latitude and/or the Solar vicinity 
(as will be shown later).
In general, even more important than the SFR is the choice of the AMR
$Z(t)$ and its dispersion, that may change
the position of simulated stars in colour-colour diagrams, 
and cause a significant colour dispersion.

After these considerations, it is convenient to specify what are to 
be considered ``default'' SFR and AMR -- i.e. those used in 
Paper~I, and partially also in this work:

The SFR was assumed to be constant over the last 11 Gyr 
for the disc, and constant between 12 and 13 Gyr for the halo.

The AMR for the disc was taken from Rocha-Pinto et al. (2000). 
[Fe/H] values are converted into the metal content $Z$ by means of a
relation that allows for $\alpha$-enhancement at decreasing [Fe/H], as
suggested by Fuhrmann (1998) data. At any age, [Fe/H] was assumed to
have a 1$\sigma$ dispersion of 0.2 dex.

The metallicity of the halo stars was assumed be $Z = 0.0095$, with a
dispersion of 1.0 dex. This was based on an observed [Fe/H] value of
$-1.6 \pm 1.0$ (Henry \& Worthey 1999), allowing for an
$\alpha$-enhancement of 0.3 dex.

\subsection{The intrinsic colour-magnitude diagrams and 
luminosity functions}
\label{sec_intrinsic}

At this point, having described the default SFR, AMR, IMF, 
and libraries of evolutionary tracks and spectra, it is useful
to open a parenthesis and illustrate the intrinsic CMDs and 
LFs that we derive from these ingredients. 
Notice that these intrinsic data are {\em the only} 
stellar ingredients that enter in the old 
``empirical $\phi(M)$ approach'' mentioned in Sect.~\ref{sec_history}.

In Fig.~\ref{fig_cmd} we show, for both disc and halo,
the $M_V$ vs. \bv\ diagram that follow from our choices,
together with the intrinsic distributions of $M_V$ and \bv\
values.

Several aspects of this figure are remarkable: 
\bite
\item It is evident that we have a very complete sampling of the 
possible evolutionary stages of the stellar populations required 
in building a realistic Galactic model.
Limiting to the bright part of the diagrams, we can notice the
presence of the young disc main sequence, subgiants and giants, 
red clump and horizontal branch stars, that make most of the 
stars observed in shallow photometric surveys. 
\item The inclusion of white dwarfs, very low-mass stars
and brown dwarfs allows us to simulate all faint stellar 
objects, including the extremely dim ones -- as dim as 
$M_V\sim44$. These faint objects are expected to appear in 
significant numbers in very deep surveys.
\item One should keep in mind that 
we know, for each simulated star, many other physical parameters
like the complete photometry, age, metallicity, surface gravity, 
etc., so that the same kind of plot could be constructed in a 
multitude of other ways and dissected by grouping stars in 
different parameter bins.
\eite

In summary, the present plots are an evindence of the significant
advantages of an evolutionary synthesis tool, over the empirical 
approaches which were used in the past to construct similar
data for Galaxy models. 

	\begin{figure}
	\resizebox{\hsize}{!}{\includegraphics{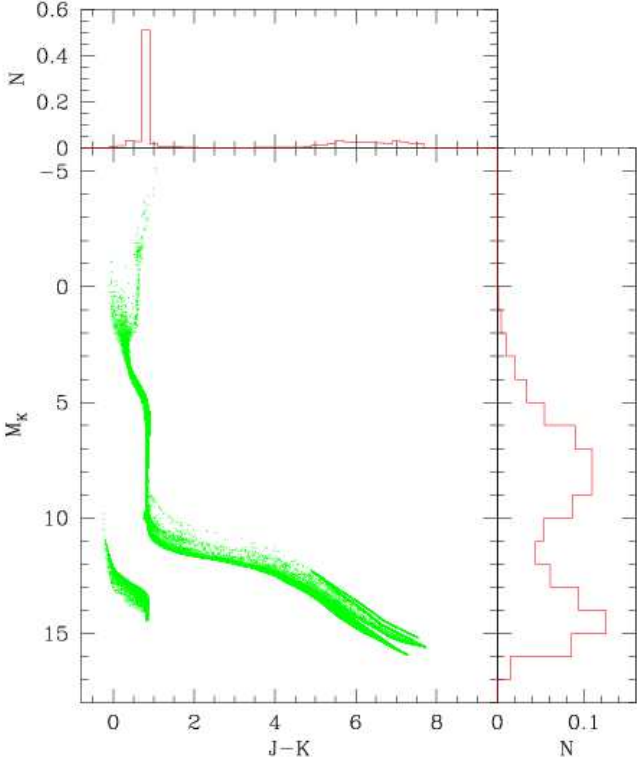}}
	\caption{The same as Fig.~\protect\ref{fig_cmd}
but limited to the disc intrinsic $M_K$ vs. \jk\ diagram.}
	\label{fig_cmdjk}
	\end{figure}

Let us also briefly comment on the general aspect of the
CMDs shown in Fig.~\ref{fig_cmd}: In the case of the disc population
shown in the left panel,
it is evident that the low-mass and brown dwarf models, taken
from different sources, combine in a continuous and well-behaved
way with the sequence drawn by more massive stellar models. 
There is just one abrupt change in the width of this main sequence,
occurring at $\mv\sim19$, that is caused by the fact that below
this limit we rely on solar-metallicity models, whereas
above it the metallicity dispersion is fully represented in the
models. Exactly the same problem is present 
for the halo population shown in the right panel. Anyway, this seems
a minor problem because -- as will be shown in the following sections --
these stars, although always present, do not make the main features
of CMDs observed up to now. Also, this seems an acceptable price to 
pay for having an extremely complete intrinsic CMD.

Finally, we recall that star counts of dwarfs below $\mv\sim6$ 
are affected just by the particular choice of IMF, whereas above 
this limit also the SFR and AMR play a major role. 

For the sake of illustration, Fig.~\ref{fig_cmdjk} shows
the intrinsic $M_K$ vs. \jk\ diagram. This looks very 
different from the former $BV$-diagram, and is 
particularly useful for the discussion of 2MASS data 
(see Sect.~\ref{sec_2mass}, and Marigo et al. 2003).

\subsection{Geometry of Galaxy components}
\label{sec_geometry}

Five are the Galaxy components presently defined in TRILEGAL:
the thin and thick discs, the halo, the bulge, and the disc
extinction layer. There is also the possibility of simulating
additional objects of known distance.

\paragraph{The thin disc:} Its density is assumed to decrease 
exponentially with the galactocentric radius projected onto the
plane of the disc, $R$,
\beq
\rho_{\rm d} = C_{\rm d} \, \exp(-R/h_R) \, f(z) \,\,,
\eeq
The vertical distribution $f(z)$ is either an exponential, 
${\rm exp}(-|z|/h_z )$, or a squared hyperbolic secant function, 
${\rm sech}^2(2\,z/h_z)$. 
Importantly, the vertical scale heigth $h_z$ is assumed to increase
with the stellar age $t$ according to the formula suggested by 
Rana \& Basu (1992):
\beq
h_{\rm d}(t) = z_0\,\left(1+t/t_0\right)^\alpha    \,,
\label{eq_hdt}
\eeq 
where $z_0$, $t_0$ and $\alpha$ are adjustable parameters.
This means that stars are formed very close to the Galaxy Plane, 
with a scale heigth $z_0$, and later disperse vertically. 
For practical reasons, the code has to deal with a limited number
of age intervals and  scale heigths. Thus, 
the total age interval is divided into a number of 
$N_{\rm d}$ subintervals -- at least 10, but typically 100 --
and the stellar densities computed separately for them.

The normalization constant $C_{\rm d}$ is set so as to produce a given
``total surface density of thin disc stars ever formed in the 
Solar Neighbourhood'', 
\[
\Sigma_{\rm d}(\odot) = \int_0^{t_{\rm G}} \psi_{\rm d}(t) \diff t 
	\int_{-\infty}^{+\infty} 
\rho_{\rm d}(\mathbf{r})_{R=R_\odot}\diff z\,\,\,,
\]
where $\psi_{\rm d}(t)$ is the SFR per unit disc area in the
so-called Solar Cylinder\footnote{The Solar Cylinder refers to a
cylinder perpendicular to the Galaxy Plane, of small diameter and 
infinite height, centered on the Sun.}, and $t_{\rm G}$ is the
Galaxy age.
So, the thin disc geometry is completely defined by the parameters 
$\Sigma_{\rm d}(\odot)$, $h_{\rm R}$, $z_0$, $t_0$ and $\alpha$.
Additionally, the parameter $N_{\rm d}$ may be adjusted to provide 
better accuracy at the cost of larger CPU times. In present
applications we use $N_{\rm d}$ as large as 100, which in fact 
represents an excellent accuracy. 

\paragraph{The thick disc:} Similarly to the thin disc, it is described 
either by a double exponential or by an exponential times a 
sech$^2$ function. However, the scale height is assumed to be 
independent of age. This because the thick disc is always described by 
predominantly old ($t\ge10$~Gyr) populations.

So, just the parameters $\Sigma_{\rm td}(\odot)$, $h_{R,{\rm td}}$, 
and $h_{z,{\rm td}}$, defined similarly to the thin disc ones, 
would suffice to define the thick disc. Needless to
say, this galactic component can be incorporated in the formula for
the thin disc, by assuming suitable scale heights $h_{\rm z}$
at large ages in Eq.~(\ref{eq_hdt}).
 
\paragraph{The halo:} Its density $\rho_{\rm h}$ is
given by a de Vaucoulers (1959) 
$r^{1/4}$ law, deprojected according to Young (1976).
Alternatively, an oblate spheroid (Gilmore 1984) can be assumed.
The halo parameters are: the radial scale $r_{\rm h}$, the
oblateness $b_{\rm h}$, and the local  
``total density of halo stars ever formed in the 
Solar Neighbourhood'', 
\[
\Omega_{\rm h}(\odot) = \rho_{\rm h}(\mathbf{r}_\odot)
	\int_0^{t_{\rm G}} \psi_{\rm h}(t) \diff t 
	 \,\,\,.
\]

\paragraph{The bulge:} Although it is included in the code
as a triaxial truncated spheroid, it has not yet been calibrated. 
For this reason no bulge component is included in any of the 
calculations presented in this paper.

\paragraph{The disc extinction layer:}
It is assumed to have an exponentially-decreasing density
in the vertical direction, with a scale-height 
$h^{\rm dust}_{z}$ (Parenago 1945, cf. M\'endez \&
van Altena 1998). The increase of $A_V$ with distance
is proportional to this density. This geometric distribution is 
normalized in two possible ways: either adopting a local
absorption density $A_V^0$, of about $0.75$ mag/kpc 
(Lyng\aa\ 1982), or adopting a total absorption at infinity 
$A_V^\infty$ as given by Schlegel et al.~(1998) maps. We set
the second option to be the default one. 
We adopt $h^{\rm dust}_{z}=110$~pc as Lyng\aa\ (1982).

Extinction is always specified in terms of $A_V$ 
($V$ in this case stands for the effective wavelength 
of Johnson's $V$-band, 5550~\AA). 
The $A_V$ values found for individual stars 
are later converted to those in the several pass-bands, $A_\lambda$, 
using the $A_\lambda/A_V$ ratios previously tabulated 
(Sect.~\ref{sec_bc}).

\paragraph{Additional objects:}
They can be inserted at known distance, and optionally assuming a known 
foreground absorption. This alternative is useful for including 
objects such as star clusters or nearby satellite galaxies in the
simulations. Of course, specifying the SFR and AMR is also
necessary, which for star clusters is limited to a single age and
metallicity value. The object total mass is again expressed in
units of total numbers of stars ever born in that area of the
sky. 

This option was recently used by Marigo et al. (2003) to simulate
the 2MASS data towards the Large Magellanic Cloud (HMC), 
and by Carraro et al. (2002) to simulate the field of the 
open cluster NGC~2158. 
Open cluster simulations in EIS fields will also be the subject of 
an upcoming paper (Hatziminaoglou et al., in preparation).

\subsection{Additional parameters}

We also have ``pointing parameters'', that specify the region of the
Galaxy sampled during a simulation. Two main modes are presently 
allowed: either (1) simulations of a projected (conic) region 
of the sky, that requires the specification of the central 
Galactic coordinates $(\ell,b)$ and total sky area, or 
(2) a volume-limited sample centered on the Sun and complete 
up to a specified maximum distance.

In both of the above cases, by specifying a given limiting 
magnitude in any of the available filters,
we avoid generating too many faint stars in the output. 

Another input parameter is the resolution in magnitudes, $\Delta m$. 
It represents the largest sub-step for the numerical integration
of Eq. (\ref{eq_counts}).
Any detail of the Galaxy geometry that is caused by a depth structure 
(in distance modulus) smaller than this resolution, or by a LF 
structure finer than it, will be lost. A resolution of 
0.1~mag is adequate for the purpose of this paper. 

The Sun's position with respect to the Galactic Plane is specified by
the galactocentric radius and height on disc.

\section{The initial calibration}
\label{sec_calib}

The initial calibration is described in Paper~I. 
The most important points, relevant for the present paper,
are repeated here. That paper describes, amongst other things, the
first application of the TRILEGAL code. The initial calibration is
derived from the six fields at high galactic latitude covered by the
``Deep Multicolor Survey'' (DMS; Osmer et al. 1998, and references
therein), and EIS data for the South Galactic Pole (SGP; Prandoni et
al. 1999). Then, the code, with the parameters fixed, was applied to
the EIS data in the Chandra Deep Field South (CDFS; Arnouts et al. 2001,
Vandame et al. 2001).

\subsection{Paper I calibration}

The IMF, SFR, and AMR for the disc and halo were those
already specified in Sects.~\ref{sec_imf} and \ref{sec_sframr}
as being the default ones. 

The disc component was described by a double-exponential in 
scale height and Galactocentric distance.  The model did not have
separate components representing the thin or thick disc. 
Instead the scale height
for disc stars was a function of age, and was parametrized as
in Eq.~(\ref{eq_hdt}). The parameter values in this equation were not
the same as in Rana \& Basu (1992) -- namely $z_0=95$~pc, $t_0=0.5$~Gyr 
and $\alpha=(2/3)$, since this does not fit very well the derived scale 
height of ``thick'', ``old'', ``intermediate'' and ``young'' 
disc components as 
derived by Ng et al.\ (1997). Their results are described by 
$z_0=95$~pc, $t_0=4.4$~Gyr and $\alpha=1.66$, which was adopted in 
Paper~I.

Since none of the six DMS fields, nor the CDFS and SGP, contains a
bulge component, this population was not included.

The Sun was assumed to be 15~pc above the Galactic Plane (Cohen 1995;
Ng et al. 1997; Binney et al. 1997) and the distance of the Sun to the
Galactic Centre was assumed to be 8.5~kpc.

With these input ingredients fixed, the halo oblateness (and local
halo number density) was derived by fitting the number of halo stars,
defined by (in the Johnson-Cousins system) $0 < \bv < 0.7$ in the
range $20 < B < 22$ and $0 < \vi < 0.8$ in the range $18 < I < 20$, in
these seven fields and was found to be $q = 0.65 \pm 0.05$. This value
was smaller than the value of $0.8 \pm 0.05$ quoted by Reid \&
Majewski (1993), but Robin et al. (2000) could not exclude a spheroid
with a flattening as small as $q = 0.6$ and Chen et al. (2001) derived
$q = 0.55 \pm 0.06$.

The disc radial scale length (and local disc number density) was
derived by fitting the number of disc stars, defined by $1.3 < \bv <
2.0$ in the range $20 < B < 22$ and $1.8 < \vi < 4.0$ in the range $18
< I < 20$, and was found to be $h_{\rm R} = 2800 \pm 250$~pc. This was
in agreement with the lower limit of 2.5~kpc (Bahcall \& Soneira 1984)
and the work of Zheng et al. (2001) on M-dwarfs who derived $h_{\rm R}
= 2750 \pm 160$ pc and Ojha (2001) who derive $h_{\rm R} = 2800 \pm
300$ pc for the thin disc.

The model with these parameters was then used to estimate the 
stellar counts in the CDFS field, yielding a fairly good fit of
the $UBVRIJK$ number counts, CMDs,
and colour distributions. The model with these parameters was also
used by Marigo et al. (2003) to successfully predict the foreground
population towards the LMC in $JHK$.

	\begin{figure}
	\resizebox{\hsize}{!}{\includegraphics{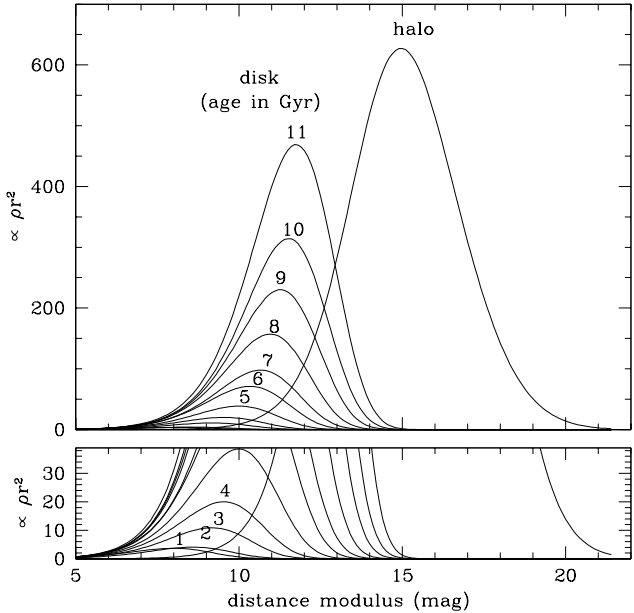}}
	\caption{Distribution of stellar distance moduli in the
simulation corresponding to our initial calibration, 
in a conic bean towards the NGP.
This is shown for 11 disc components of increasing age -- 
at steps of 1 Gyr, as labelled -- and for the halo. 
The top panel shows all the curves, whereas the 
bottom one expands the vertical axis in order
to detail the profiles for ages younger than 5 Gyr.
It can be noticed that younger disc components
are found at lower mean distances (peaking from say $\dmo=8$ 
to 12 as the age increase), whereas the halo stars are found with a
nearly-Gaussian distribution of $\dmo$ which peaks at about 15.
The characteristic shapes of these distributions result from 
two competing trends in Eq.~(\protect\ref{eq_counts}): 
the quadratic increase of the geometrical
factor $r^2$, and the almost-exponential decrease of the density 
$\rho$ with distance to the Galactic Plane. }
	\label{fig_profile}
	\end{figure}

\subsection{Distribution of distance moduli}

It is very instructive to look at the characteristic 
distributions of distance moduli, $\dmo=5\log r-5$, 
for this model calibration. Were all
stars in a given field -- even the dimmest
ones -- possible to be observed, such distribution would be 
proportional to the integral of the quantity $\rho\, r^2$ 
(see Eq.~\ref{eq_counts}) over small bins of distance modulus.
We show these quantities as evaluated for the line of 
sight of the North Galactic Pole (NGP) 
in Fig.~\ref{fig_profile}. In the case of the
disc, we separate the profiles coming from different ages (i.e. 
different scale heights) spaced by 1 Gyr. As can be noticed, 
a simulation of the NGP -- if not constrained by 
any limiting magnitude -- would contain increasing numbers of 
disc stars as we go to older ages, 
and at increasing mean distances (from $\dmo\simeq6$ to 12 as the 
age goes from very young to 11~Gyr). For each age considered,
the disc distribution of \dmo\ looks like an asymmetric 
curve with a slow increase followed by a faster decay. 
The \dmo\ distribution of halo stars, instead, 
looks like a single Gaussian of mean $\dmo\simeq15$.

As a rule, we can conclude that halo stars dominate star counts at
very large distance moduli ($\dmo\ga13$), whereas intermediate-age to
old disc stars would dominate counts at most ``intermediate distances'' 
($\dmo$ from say 9 to 13). Only at very short distance moduli -- 
$\dmo\la9$, i.e. in the immediate Solar Neighbourhood -- 
can the young disc stars make a sizeable contribution to the 
star counts.
Alternatively, one has to look at lower galactic latitudes to see
a higher contribution from the young disc.

Of course, the situation gets more complex as we consider the
limiting magnitudes that are present in any survey, and that favour
the detection of the few closest stars in spite of the many 
distant ones. Anyway, the present Fig.~\ref{fig_profile} shows the
type of stars which make the major contribution depending on the
depth of a given survey. This information is relevant for the 
discussion presented in the next section.

\section{Recalibration and fine tuning}
\label{sec_finecalib}

Since Paper~I, we have improved many aspects of TRILEGAL,   
and checked the model predictions with additional datasets.
This has forced some changes in the model calibration, 
as will be detailed in the subsections below. 

	\begin{figure*}
	\begin{minipage}{0.46\textwidth}
	\resizebox{\hsize}{!}{\includegraphics{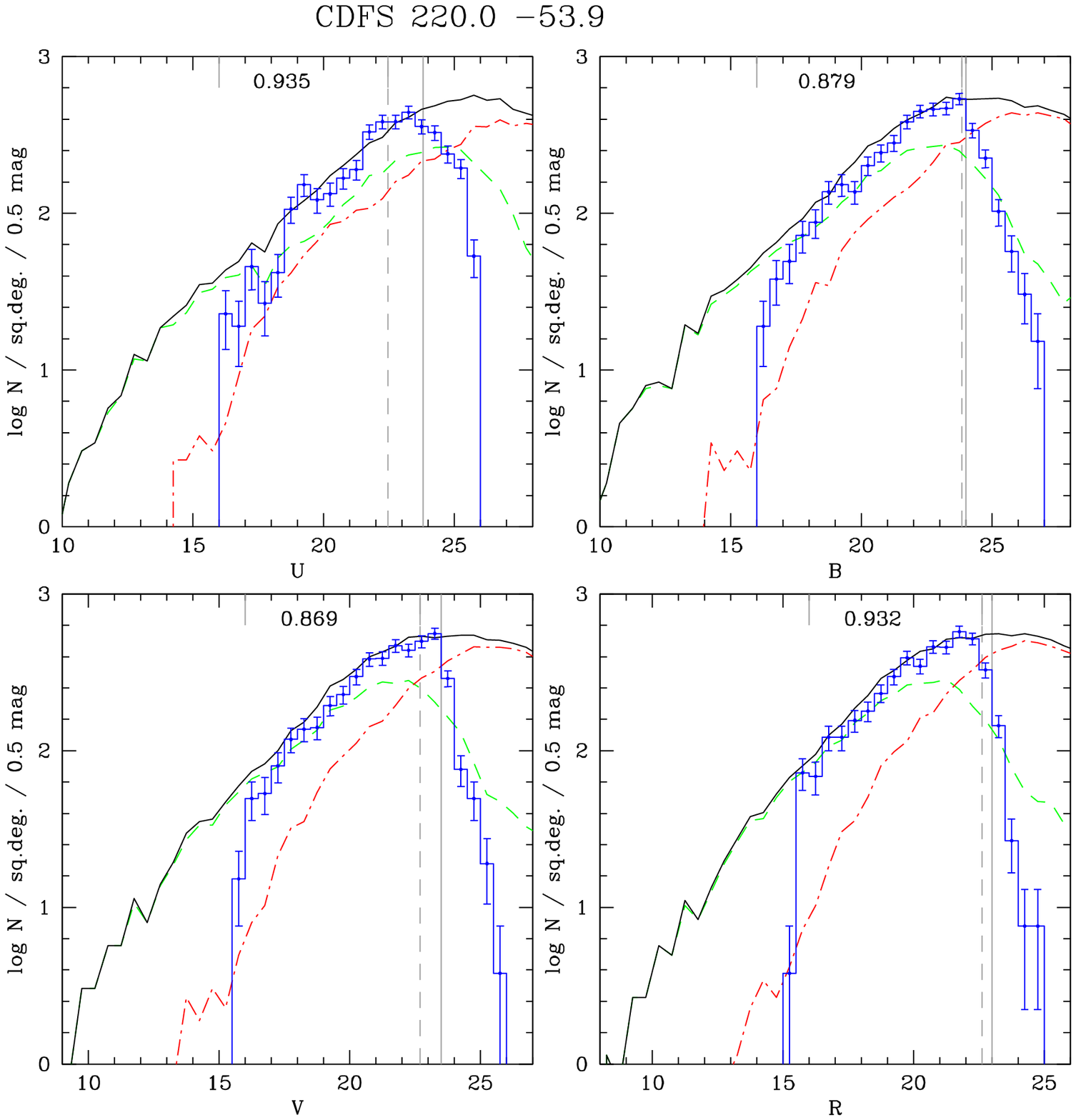}}
	\end{minipage}
	\hfill
	\begin{minipage}{0.46\textwidth}
	\resizebox{\hsize}{!}{\includegraphics{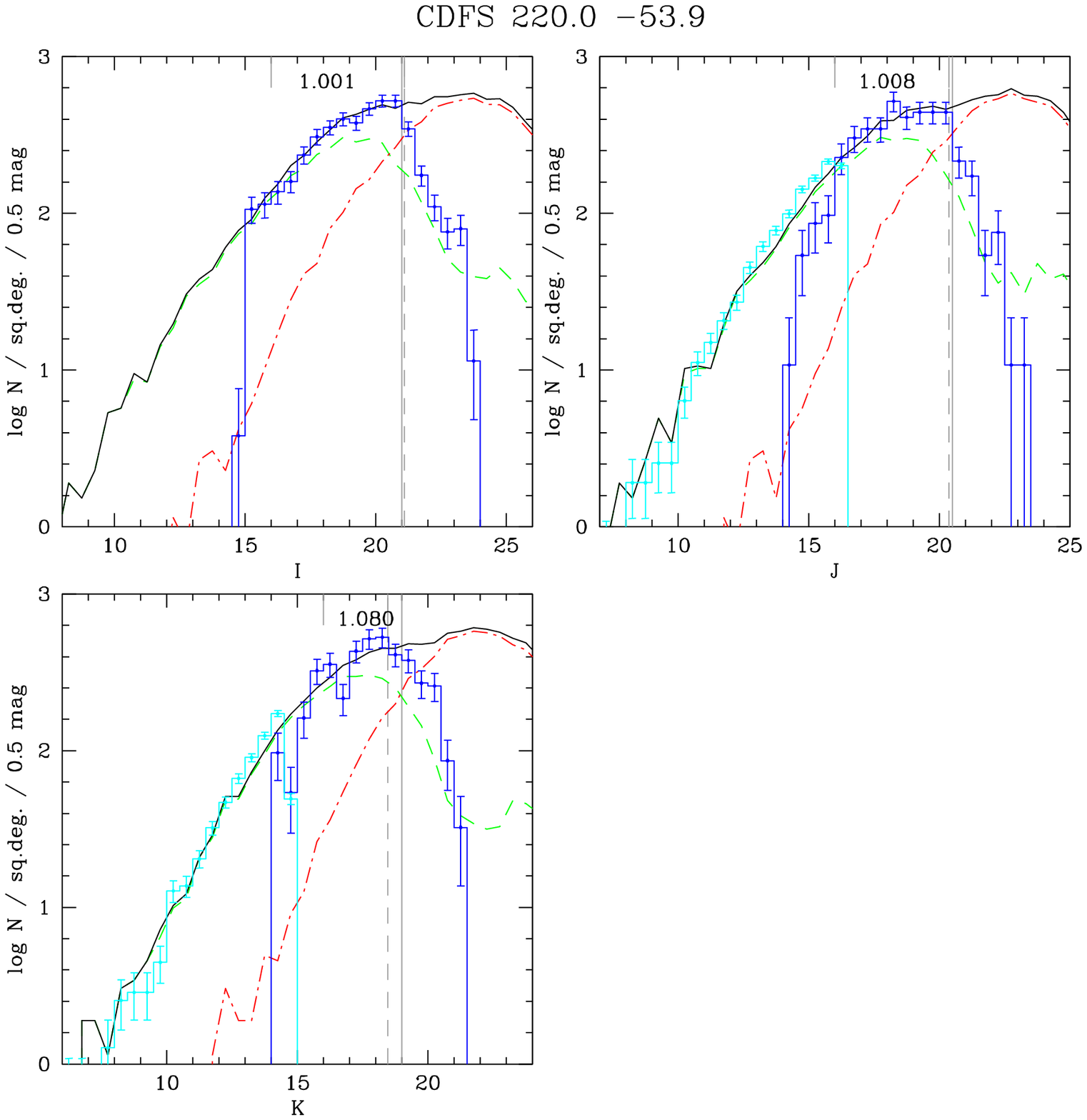}}
	\end{minipage}
	\caption{Number counts as a function of magnitude. 
The histogram with error bars represents 
the data from the 7 and 5-passband CDFS stellar 
catalogues (Paper~I) with poissonian errors. 
The dashed and continuous gray vertical lines 
indicate the magnitude limits for efficient morphological
classification, {\tt MAG\_STAR\_LIM}, and the 90 percent 
completeness limit, respectively.
For the $J$ and $K$ plots, at brighter 
magnitudes ($J<16.5$ and $K<15$, respectively) we add the
histogram corresponding to 2MASS $JK_{\rm s}$ data.
The smooth lines refer to the model results for a region of much 
larger area and hence better statistics, computed with the same
binning (0.5~mag) as the data:
they show separately the contribution of the halo (dot-dashed line), 
of the disc (dashed), and the total result (continuous line). 
At the top of the plot, we present 
the number ratio between observed and modelled sources as counted 
within the limits indicated by short gray vertical marks
(the fainter limit coincides with {\tt MAG\_STAR\_LIM}). 
Inside these magnitude limits, the comparison
between the continuous line and histogram indicates the goodness 
of the model in reproducing the observed star counts.
}
	\label{fig_cdfs}
	\end{figure*}

	\begin{figure*}
	\begin{minipage}{0.46\textwidth}
	\resizebox{\hsize}{!}{\includegraphics{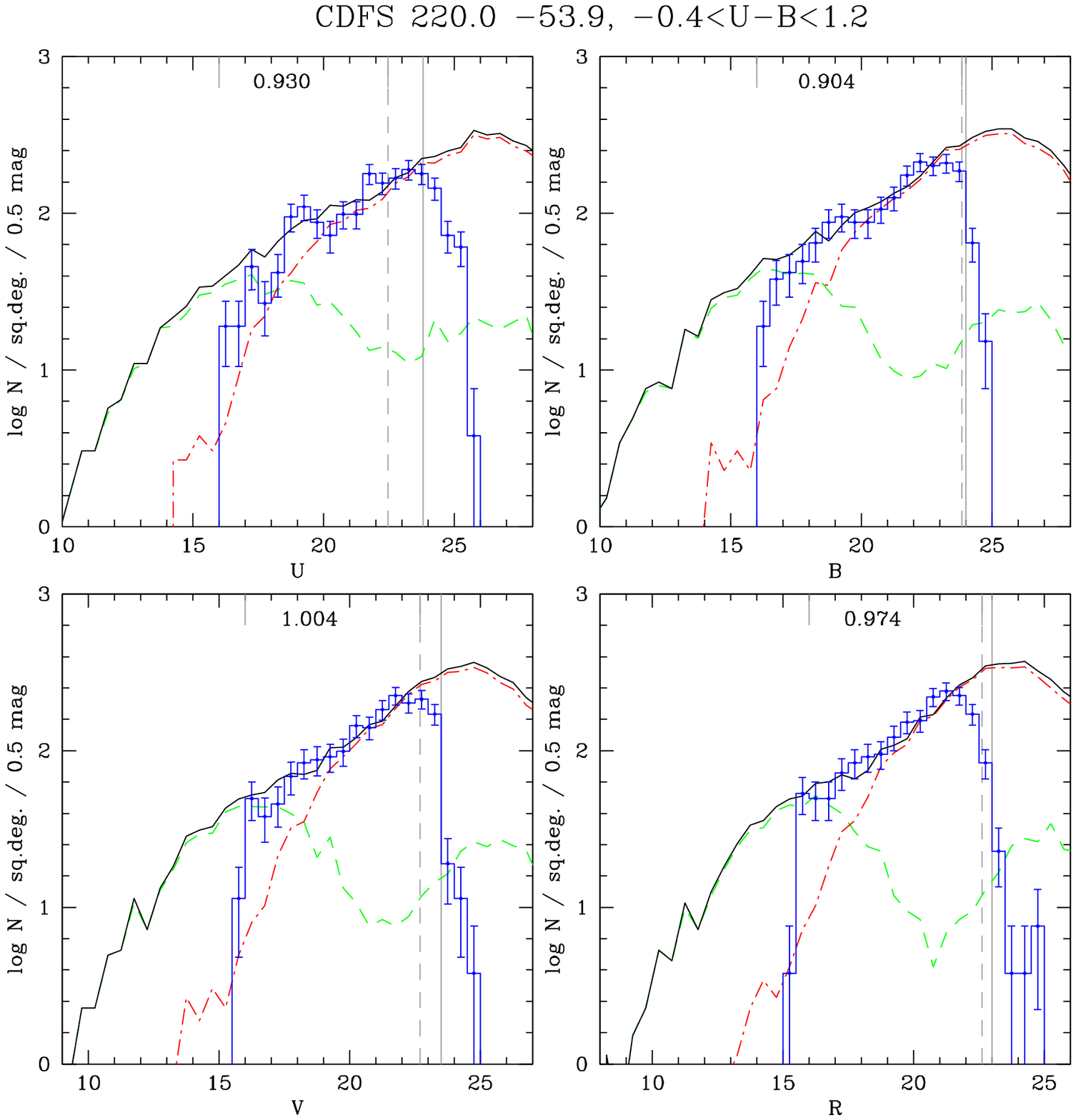}}
	\end{minipage}
	\hfill
	\begin{minipage}{0.46\textwidth}
	\resizebox{\hsize}{!}{\includegraphics{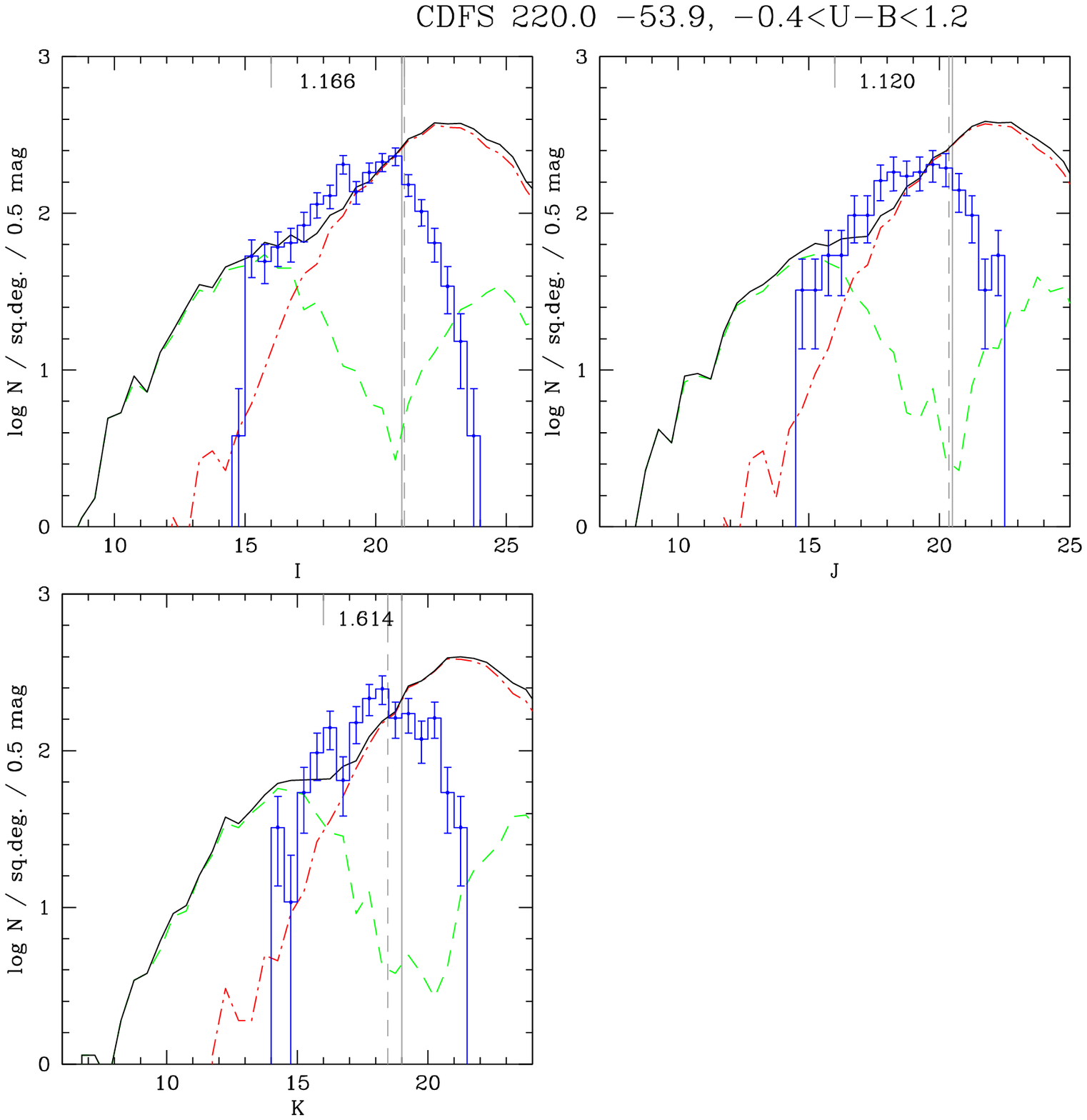}}
	\end{minipage}
	\caption{The same as Fig.~\protect\ref{fig_cdfs},
but now limiting data and models to the ``blue subsample'' 
with $-0.4\le\ub\le1.2$, which is dominated by the disc 
at brighter magnitudes and by the halo at the faintest.
To help in the comparison, the scales are kept the same
as in Fig.~\protect\ref{fig_cdfs}.
The comparison between data (histogram with error bars)
and models (continuous line), inside the magnitude interval
delimited by the vertical gray marks, probes whether the
models correctly predict the number ratio between 
halo and disc stars (shown separately by dot-dashed and 
dashed lines, respectively).
}
	\label{fig_cdfs_blue}
	\end{figure*}

\subsection{Summary of recent changes}

First of all, we opted for a better temporal accuracy as 
given by $N_{\rm d}=100$ (see Sect.~\ref{sec_geometry}) 
instead of the $N_{\rm d}=10$ adopted in 
Paper~I. $N_{\rm d}=100$ implies a virtually 
continuous change of thin disc geometry with age, which
in itself represents a novelty in star count models.

Second, we have adopted a more realistic metallicity 
distribution for halo stars, namely the one measured by
Ryan \& Norris (1991).
The Sun's height above the disc was corrected to
$24.2$~pc (cf. Ma\'\i z-Apell\'aniz 2001).

So far, these changes have just a minor impact in our final
number counts. Major revisions instead resulted from our
attempts to reproduce 2MASS and Hipparcos data, not considered
in Paper~I. Without
entering in a detailed description of all attempts carried out
to fit the models to the data, here we describe the main arguments
used in establishing our final calibration. 
They are:

1) The Paper~I calibration, although in
excellent agreement with deep star counts and with the 
constraints from Hipparcos, presented systematically too few
stars at intermediate magnitudes, of say $K\sim14$, as 
compared to 2MASS data. The
deficit was small for high latitudes, reaching factors of 
about 2 at $b\sim10\deg$. This calls for an decrease in the slope
of the LFs.

2) Star counts at faint magnitudes are determined essentially
by the halo plus the oldest disc components (those with
highest scale heights, see Fig.~\ref{fig_profile}). 
A way to decrease the LF slope is to 
reduce the contribution of the oldest disc components, and 
then compensate the reduction in total number counts by
increasing the total disc SFR. We perform such an operation
in the following way: the age scale of the thin disc SFR
is changed by a constant factor of 0.8 (or $-0.1$ dex in 
$\log t$), so that the oldest disc component is found at an 
age of 9 Gyr, and with scale height of $h_z=603$~pc.

3) This change in the age (and height) scales of the disc 
is quite satisfactory in predicting star counts in CDFS,
DMS, and 2MASS, but now fails by predicting twice
as many stars as sampled by Hipparcos. This local sample 
is dominated by relatively younger stars, of ages $\la2.5$ Gyr 
(as indicated by Fig.~\ref{fig_profile}) and $h_z\la150$ pc. 
In order to eliminate this discrepancy,
we simply change the vertical distribution of thin disc 
stars, from the usual exponential to a squared hyperbolic 
secant law with twice its scale height, i.e.\
\beq
	{\rm exp}(-|z|/h_z) \rightarrow 0.25\,{\rm sech}^2(0.5\,z/h_z)
\eeq 
In this way, star counts at deep and intermediate magnitudes 
do not change by much, while the local one is reduced  
to the level of the observed values.

The assumption of a sech$^2$ function instead of an exponential one is 
far from being arbitrary. Exponential laws are commonly assumed in 
star count models mainly because this is inferred from the surface 
brightness profiles in the discs of edge-on galaxies. 
But this observation does not refer to the inner part of the
discs, which are normally obscured by dust lanes and are the site of 
ongoing star-formation. 
From the kinematic point of view, theory predicts that the
density profile of isothermal discs do follow a sech$^2$ law 
-- then approaching an exponential law at high $z$, in accordance with
observations.
Of course, the disc of our Galaxy is not isothermal -- 
as demonstrated by the observed increase of scale heights 
with the stellar ages --, but this latter theoretical aspect
makes the sech$^2$ law to be a better approximation than a
simple exponential law.

Finally, in order to improve upon our final results for
2MASS and Hipparcos, 
he have modified the SFR of the thin disc: we have
assumed that between 1 and 4 Gyr the SFR has been 1.5 times
larger than at other ages. This causes a moderate impact 
in the distribution of stars in the \mv\ vs. \bv\ diagram, 
that can only be explored by means of Hipparcos data. 
In fact, such a change in the disc SFR has a negligible impact 
on deep fields and just a minor impact on 2MASS counts.

The normalization constants we derived imply
\bite
 \item a local surface density of ever-formed disc stars of 
$\Sigma_{\rm d}(\odot)=59$~$\Msun\,{\rm pc}^{-2}$, a number that
compares well with the present dynamical surface density 
of matter in the Solar Cylinder
($56\pm6$~$\Msun\,{\rm pc}^{-2}$ cf. Holmberg \& Flynn 2004). 
We find reassuring that these two quantities, being very
different in their nature, have the same order of magnitude. 
This would indicate an efficient conversion of baryonic matter 
into stars over the history of our disc.
 \item a local volume density of halo stars ever-formed
$\Omega_{\rm h}(\odot)=1.5\times10^{-4}$~$\Msun\,{\rm pc}^{-3}$.
\eite

Let us now present the results for this calibration,
providing more details on the way the different data samples 
have been selected and modelled.

\subsection{Simulating deep fields: CDFS, DMS, and SGP}
\label{sec_deep}

A main characteristic of deep photometric surveys is the 
rich presence of galaxies, a significant fraction of which appears
as point sources and cannot be easily distinguished from real
stars. Thus, object classification by means of morphological and
photometric criteria is of central importance in these fields.
The reader is referred to Paper~I and
Hatziminaoglou et al. (2002) for a discussion of these aspects.

In this paper, we deal with 3 deep catalogues which, as far as possible,
have been cleaned from contamination by galaxies.

\subsubsection{CDFS}

The first one is the CDFS stellar catalogue 
(Paper~I), which points towards a relatively 
clean area centered in $(\ell=220.0,b= -53.9)$. According to 
Schlegel et al. (1998) maps, the reddening for background sources
amounts to $\ebv=0.0148$ or $A_V=0.0458$.
We make use of $UBVRI$ data from the 5-passband catalogue 
covering 0.263~deg$^2$, and $JK$ from the
7-passband catalogue covering 0.0927~deg$^2$. 
The data has been cleaned from non-stellar objects according to the
criteria and methods thoroughly discussed in Paper~I.
For this field, we simulate the galactic population for a 2.63~deg$^2$
region. Fig.~\ref{fig_cdfs} shows the results in units of number of 
stars per unit deg$^2$ and 0.5-mag intervals. As can be noticed, the 
agreement between modelled and observed counts is good, their
ratio being comprised between 0.87 and 1.08 for all 7 pass-bands
considered.  

Particularly interesting is the comparison between simulated and
observed number counts for stars in the interval $-0.4\le\ub\le1.2$.
This subsample, according to the models, is completely dominated 
by disc stars at brighter magnitudes ($V\la18$), and then by
halo stars at fainter magnitudes ($V\ga20$); this is
illustrated by the comparisons between observed and 
simulated CMDs shown in Paper~I (see 
in particular the online version of their figure~5). 
Hence, by fitting number counts in this particular subsample we can 
be sure to be correctly modelling the relative proportion 
between halo and disc densities.
Moreover, a ``blue subsample'' defined in this way is composed
mostly by stars in well-modelled evolutionary stages -- 
i.e. main sequence stars of moderately high \Teff -- and 
excludes the hot white dwarfs and the reddest 
very-low mass stars, for which the reliability of present-day 
evolutionary and spectral models could still be questioned.

Such a comparison is presented in Fig.~\ref{fig_cdfs_blue}.
As can be noticed, the blue subsample 
presents good evidence that the ratio
between halo and disc densities is well represented
for this line-of-sight. The most significant discrepancy that 
appears at this point is a moderate excess of halo stars at
$B\simeq23$. As demonstrated in Paper~I, 
this is a magnitude interval in which 
number counts become sensitive to the very low-mass IMF.
In fact, we have verified that a better agreement with 
observations turns out if we artificially eliminate halo stars
with $\mini\la0.2$~\Msun\ from our models. However, we prefer
not to draw any strong conclusion from this test, since
$B\simeq23$ is also close to both the limit for efficient 
morphological classification, {\tt MAG\_STAR\_LIM}, and the
90-percent completeness limit of CDFS data. 

\subsubsection{DMS}

The Deep Multicolor Survey by Hall et al. (1996) and
Osmer et al. (1998) provides deep $UBVR'I75$ and $I86$ data 
for 6 different fields of $|b|\ga30\deg$.
From their catalogue, we eliminated the objects that are likely 
not to correspond to stars, i.e. those marked as
``galaxy'', ``noise'', ``diffuse object'', or ``long object''
in any of the DMS passbands.

	\begin{figure*}
	\begin{minipage}{0.44\textwidth}
	\resizebox{\hsize}{!}{\includegraphics{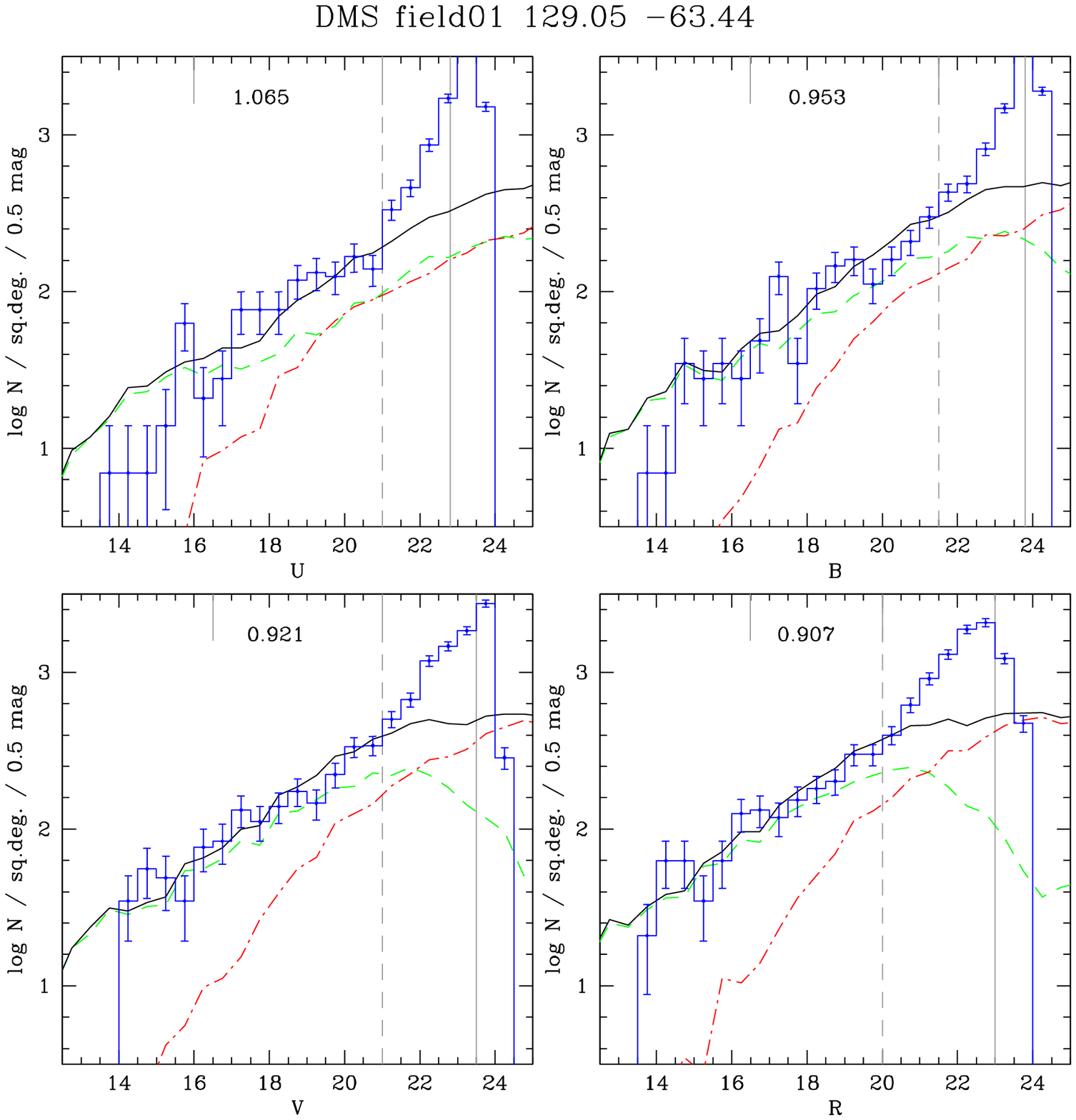}}
	\resizebox{\hsize}{!}{\includegraphics{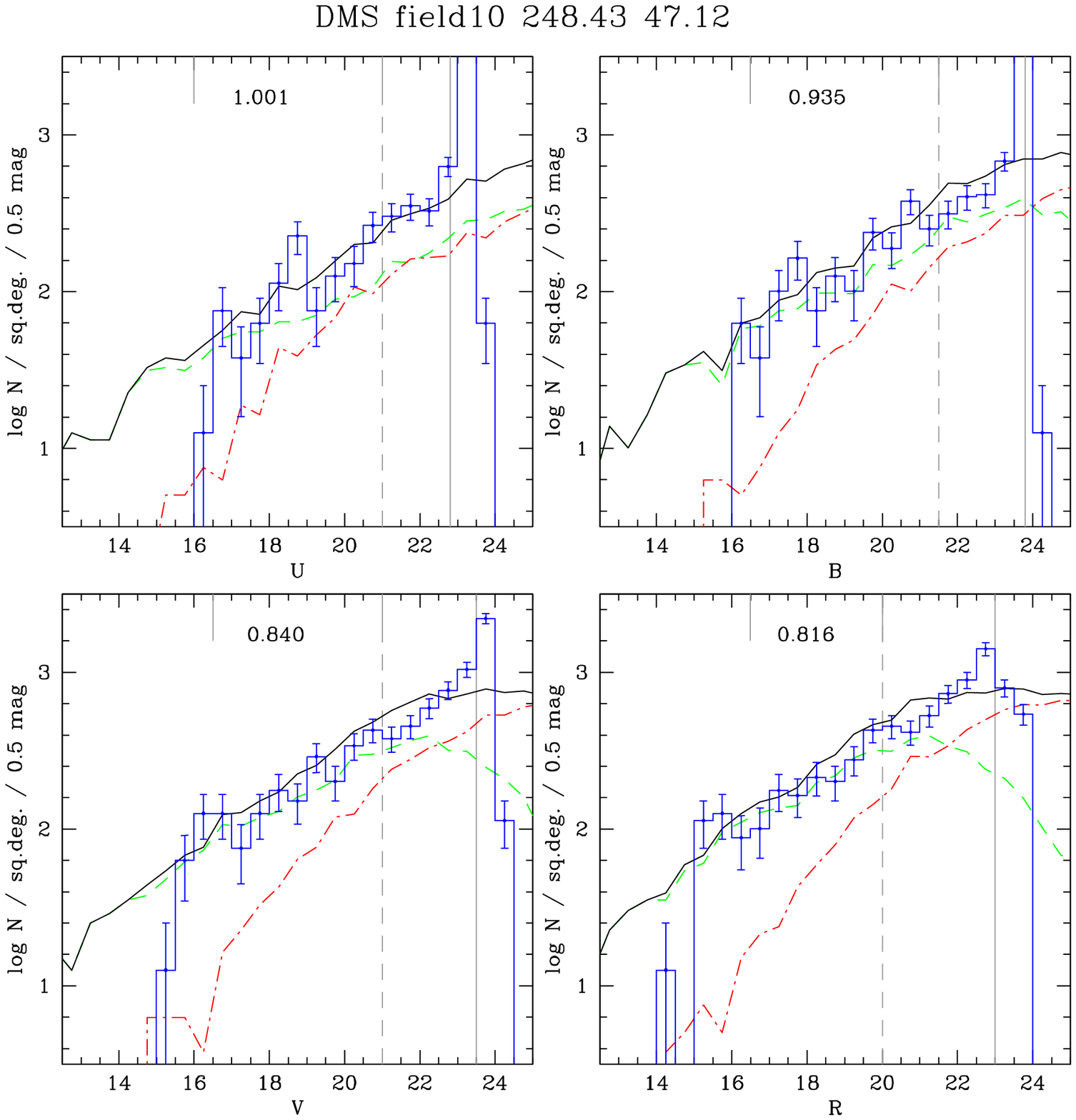}}
	\resizebox{\hsize}{!}{\includegraphics{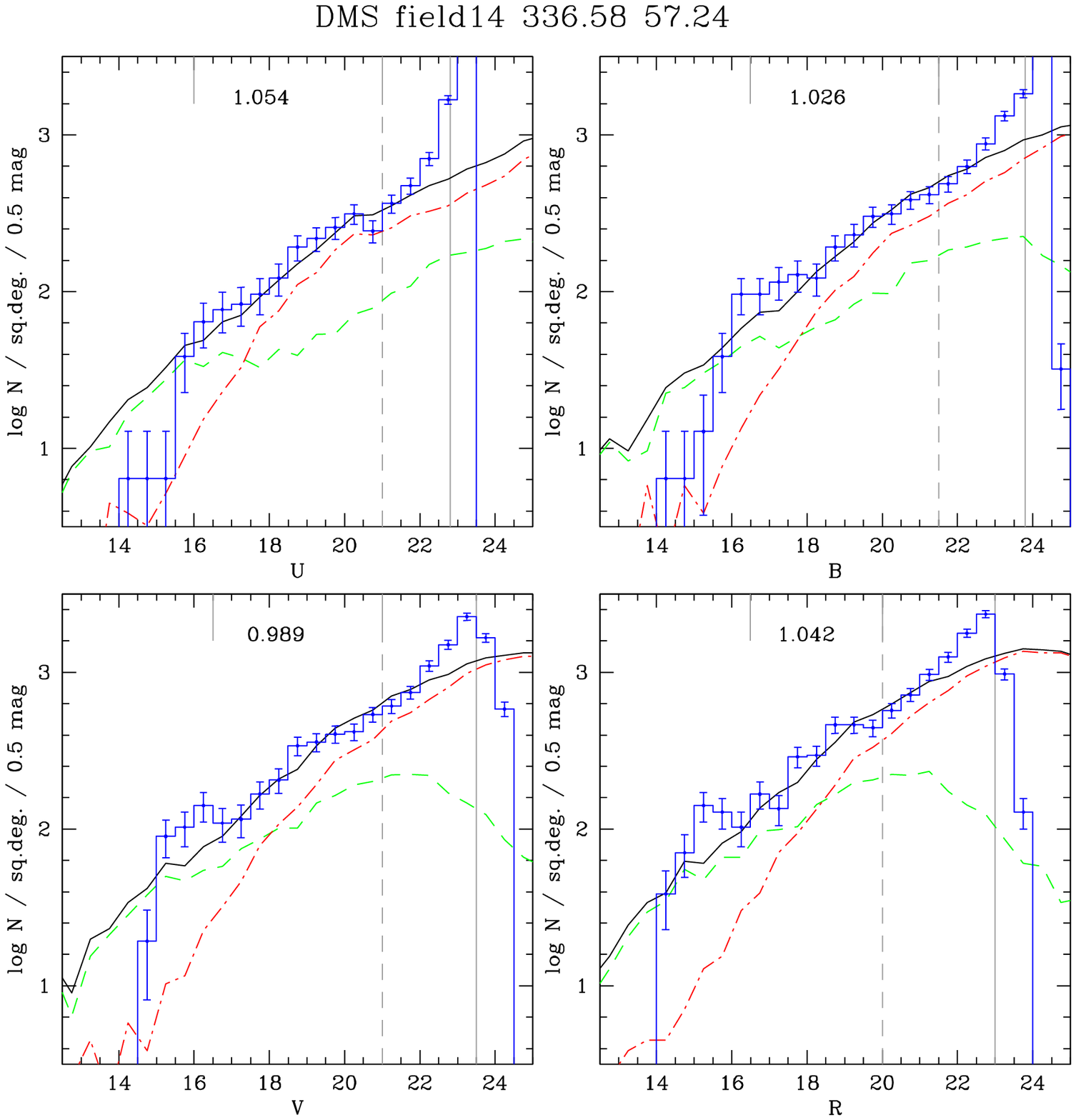}}
	\end{minipage}
	\hfill
	\begin{minipage}{0.44\textwidth}
	\resizebox{\hsize}{!}{\includegraphics{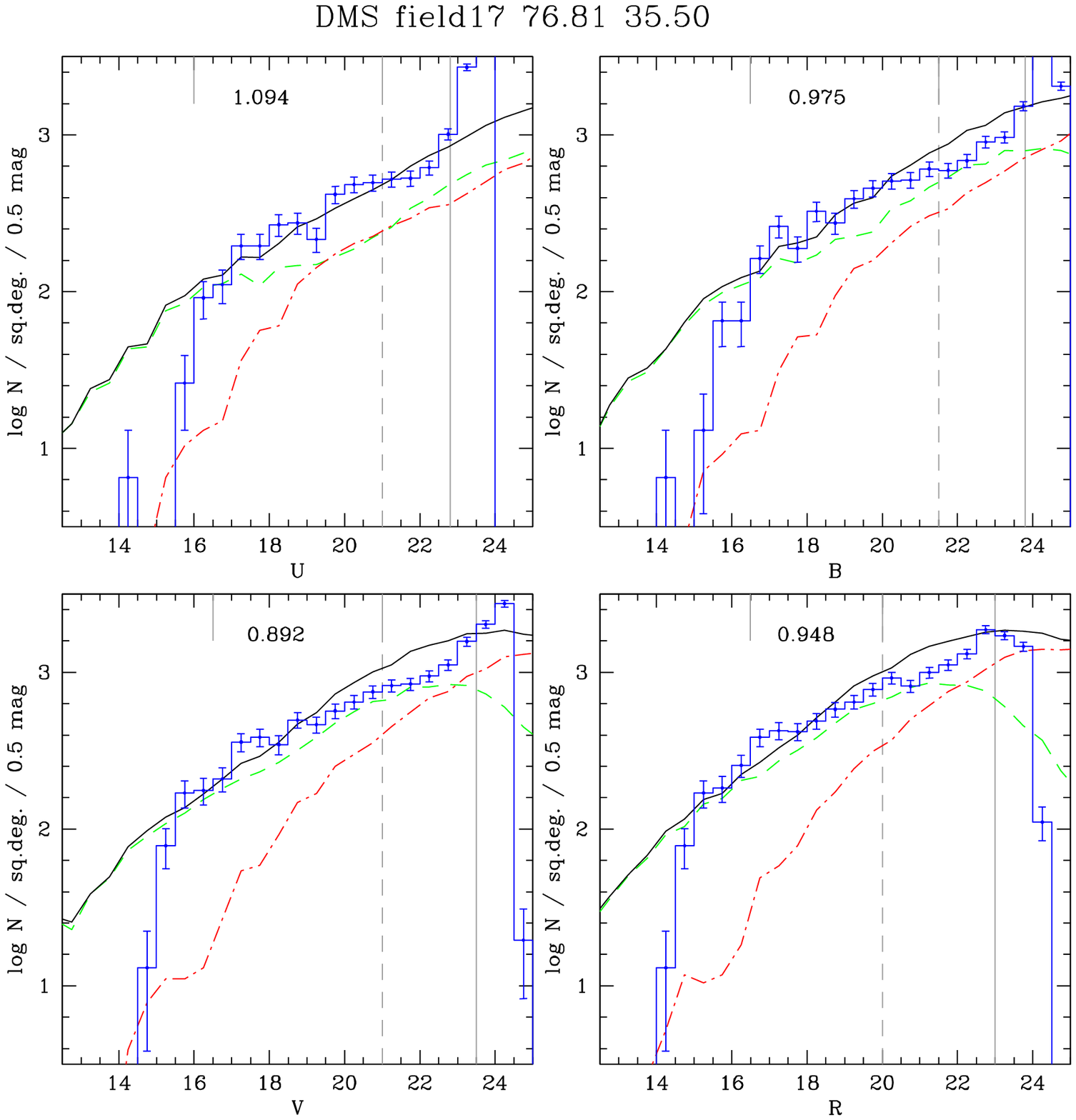}}
	\resizebox{\hsize}{!}{\includegraphics{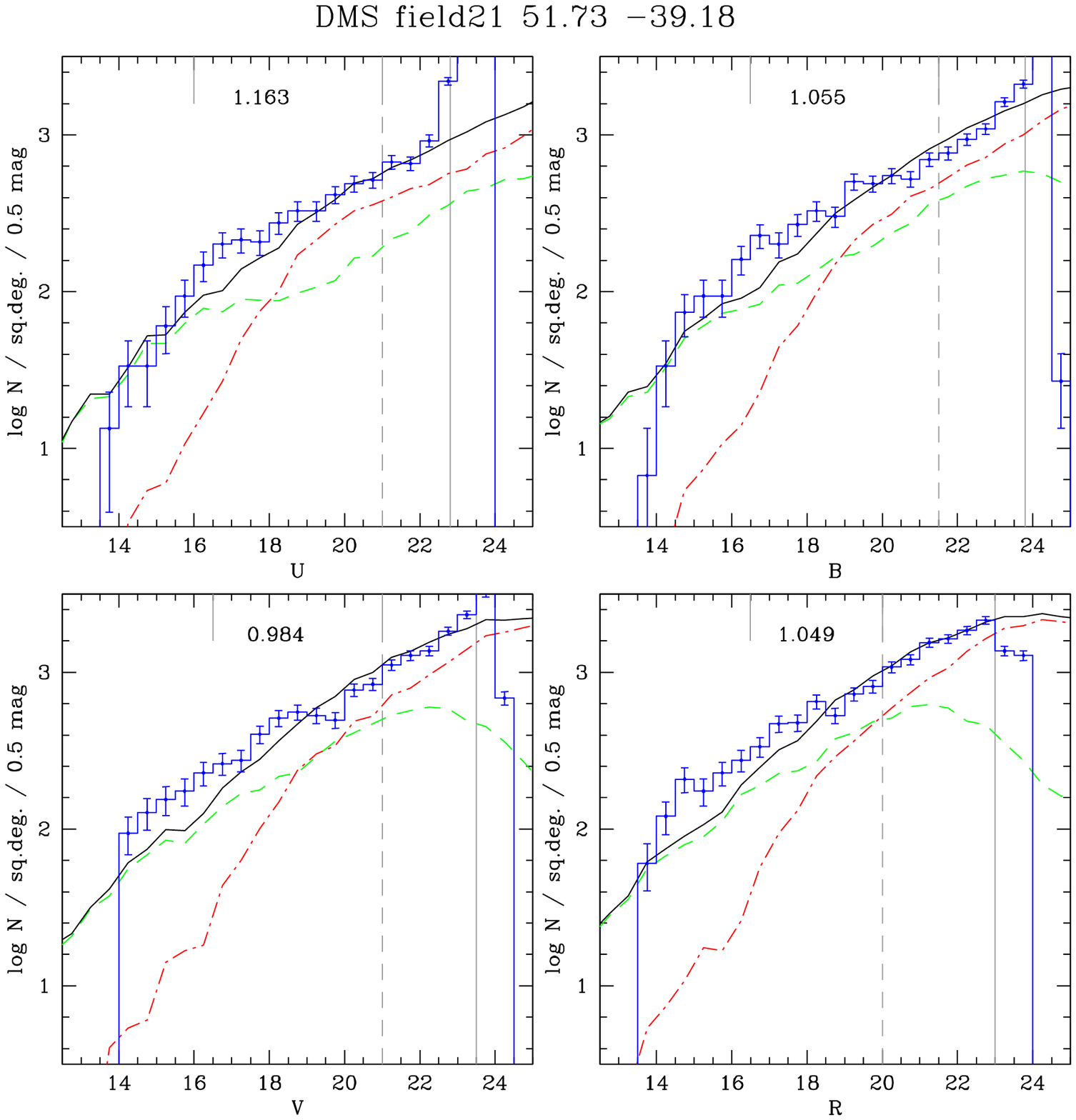}}
	\resizebox{\hsize}{!}{\includegraphics{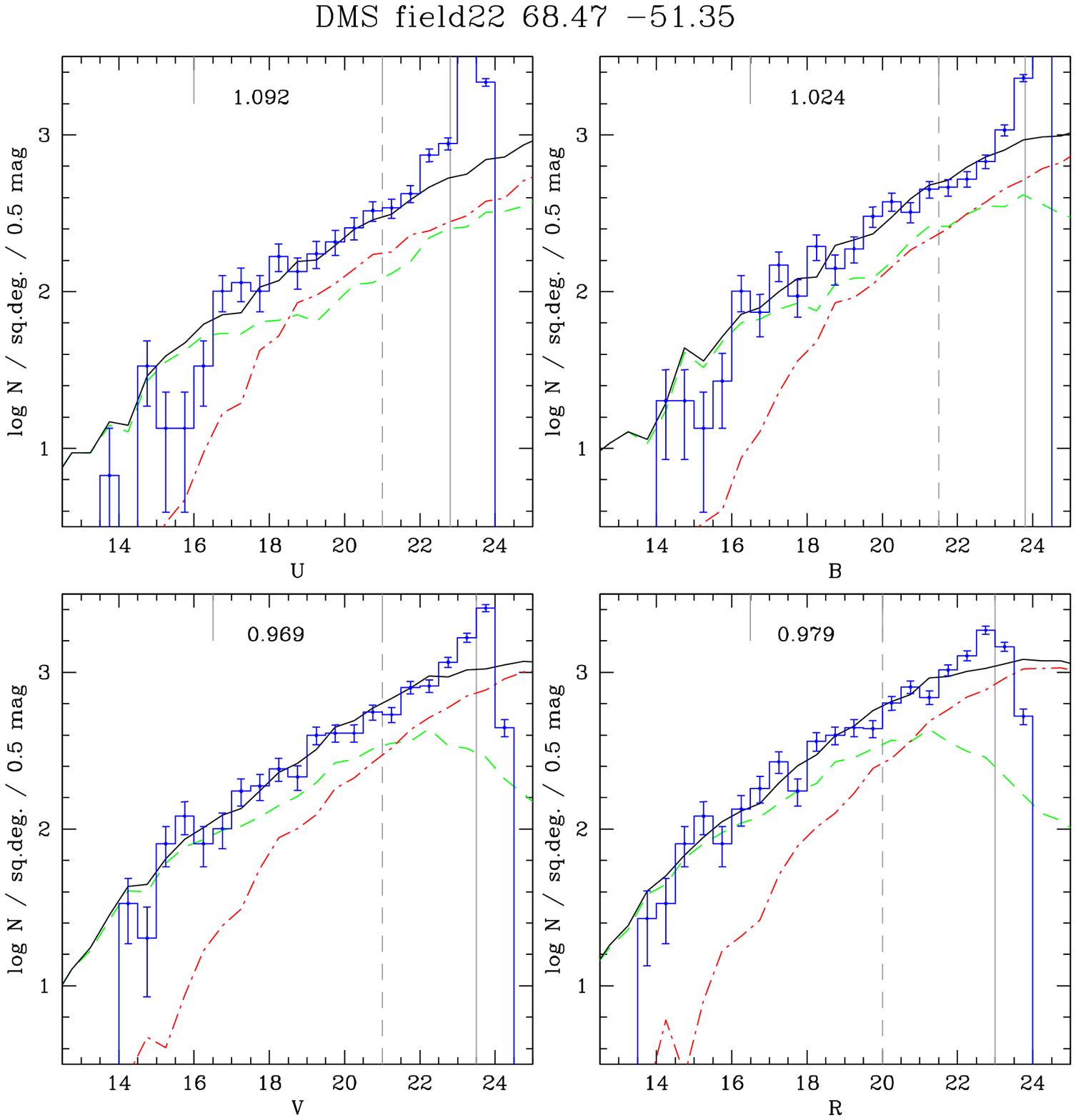}}
	\end{minipage}
	\caption{The same as Fig. \protect\ref{fig_cdfs},
but for the several fields of the DMS (Osmer et al. 1998),
whose $(\ell,b)$ are given at the top of each panel.
The data (histogram with error bars) excludes the 
objects that are likely not to correspond to stars. 
The gray vertical lines this time indicate the DMS ``threshold magnitude'' 
(dashed line) and the ``$5\,\sigma$ limiting magnitude'' (also the
$90$ percent completeness magnitude; continuous line) as 
determined by Hall et al. (1996). The star counts ratio between model 
and data (shown in the upper part of each panel) is computed between 
the ``upper limiting magnitude'' (16.0 or 16.5, depending on the
passband) and the threshold one.
}
	\label{fig_dms}
	\end{figure*}

Figure~\ref{fig_dms} shows the $UBVR$ results as compared to 
our models. 
As shown in the figure, if we limit the analysis 
to magnitudes between the DMS ``upper limiting'' and the
``threshold'' one -- an interval 
almost free from saturation, incompleteness and 
contamination effects -- there is a good overall 
agreement in the number counts between the data and our model.
With the exception of DMS field 21, the largest discrepancies 
reach about 10 percent in number counts. 

\begin{figure*}
	\begin{minipage}{0.44\textwidth}
	\resizebox{\hsize}{!}{\includegraphics{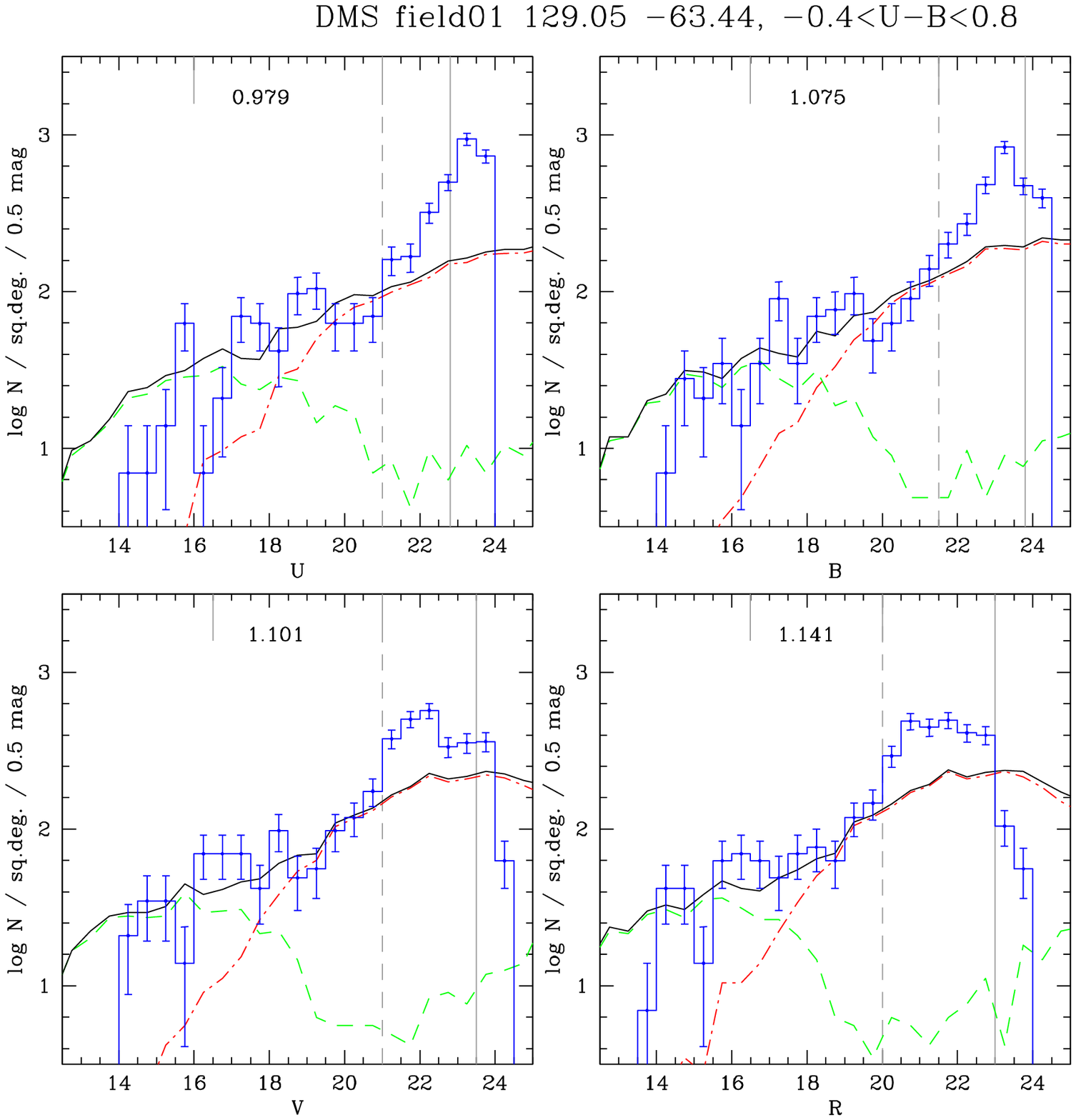}}
	\resizebox{\hsize}{!}{\includegraphics{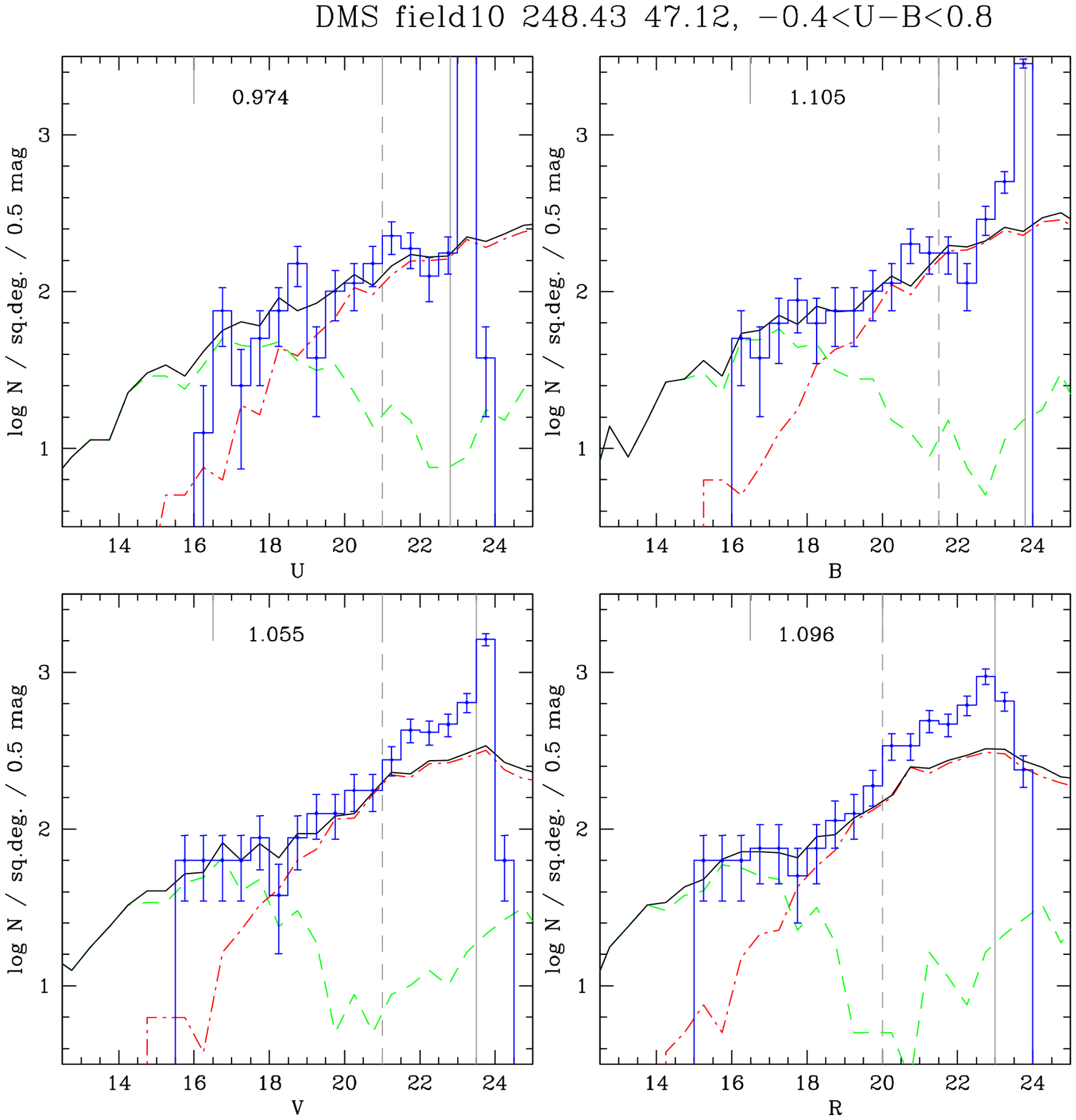}}
	\resizebox{\hsize}{!}{\includegraphics{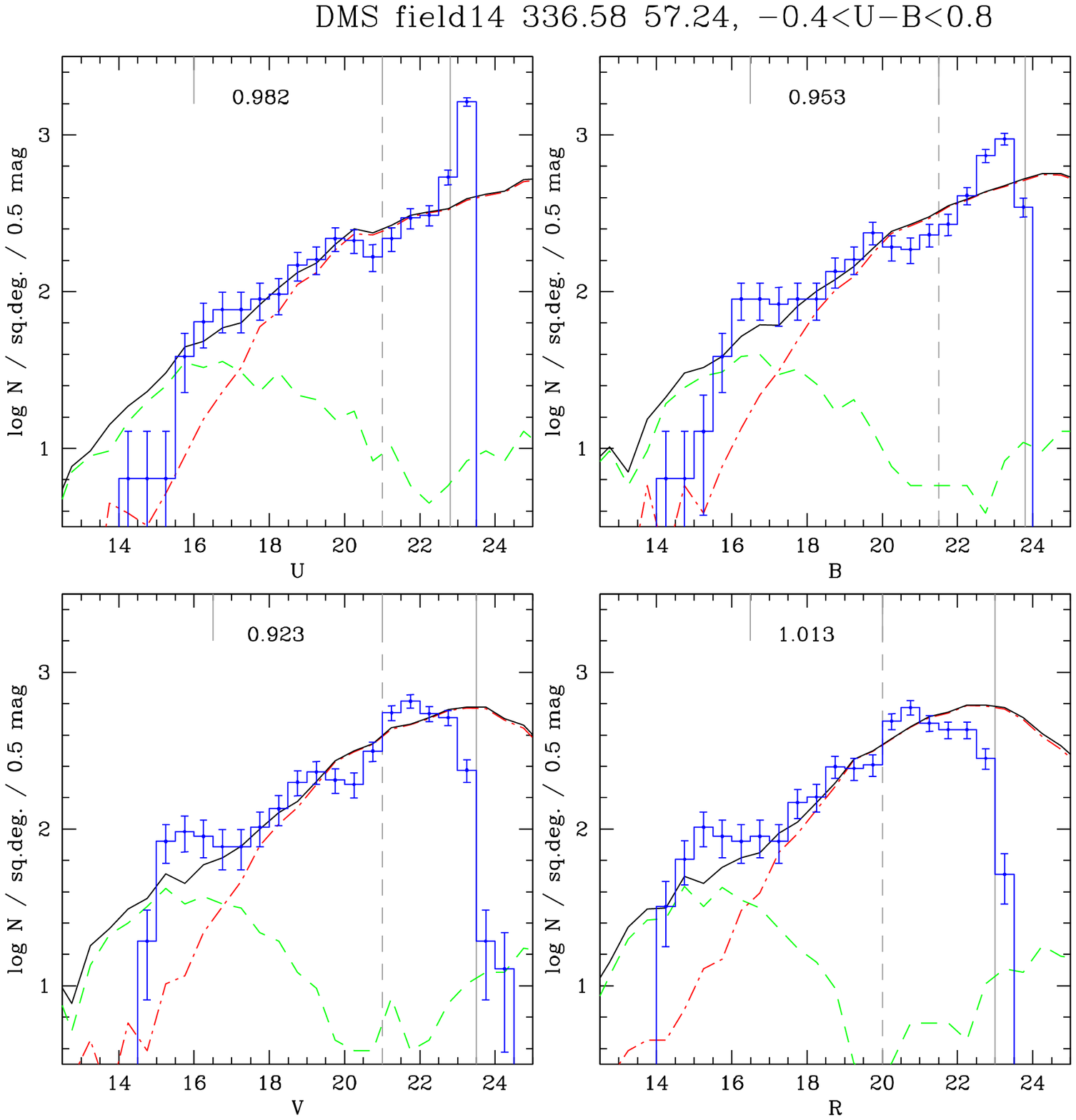}}
	\end{minipage}
	\hfill
	\begin{minipage}{0.44\textwidth}
	\resizebox{\hsize}{!}{\includegraphics{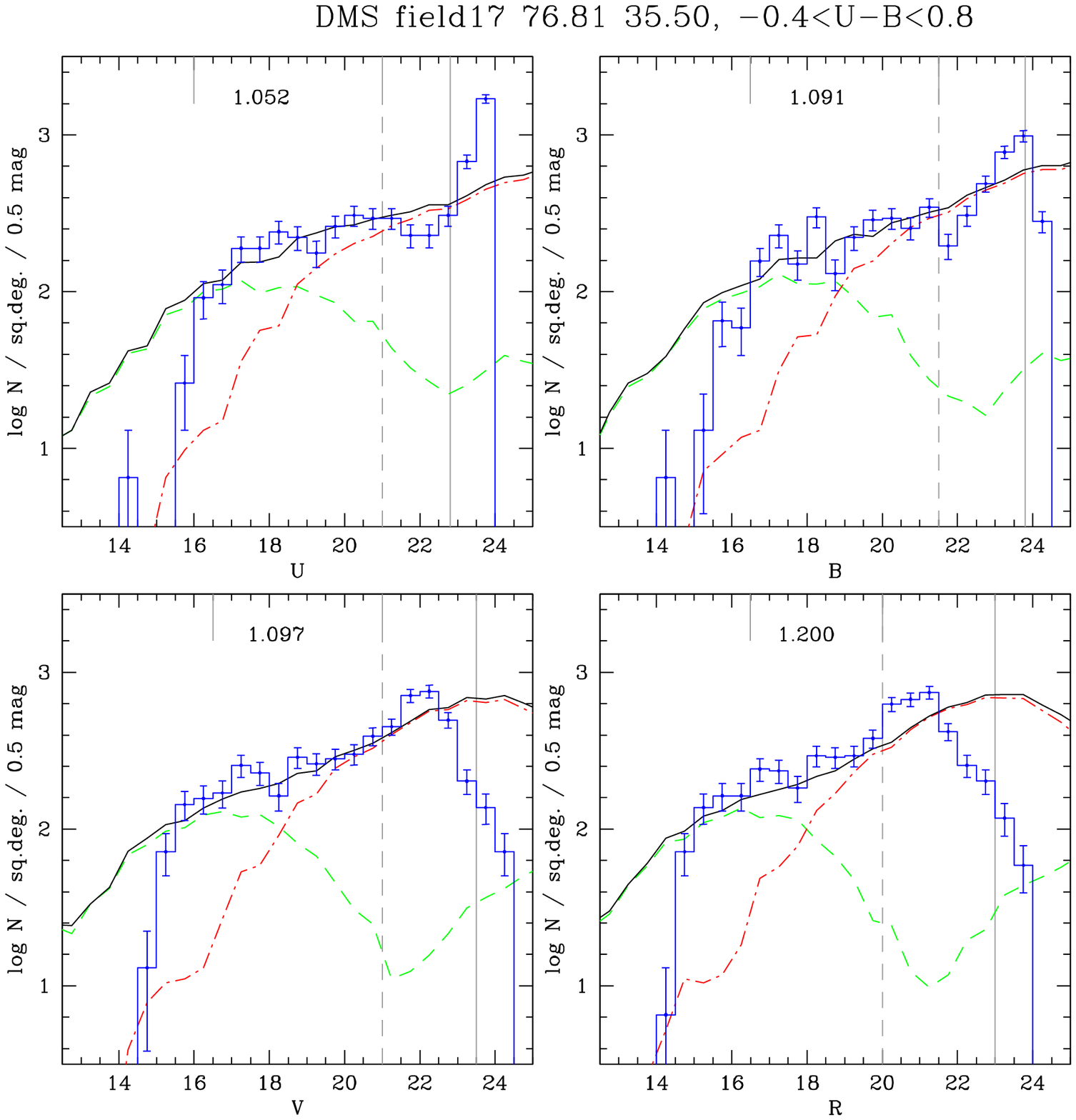}}
	\resizebox{\hsize}{!}{\includegraphics{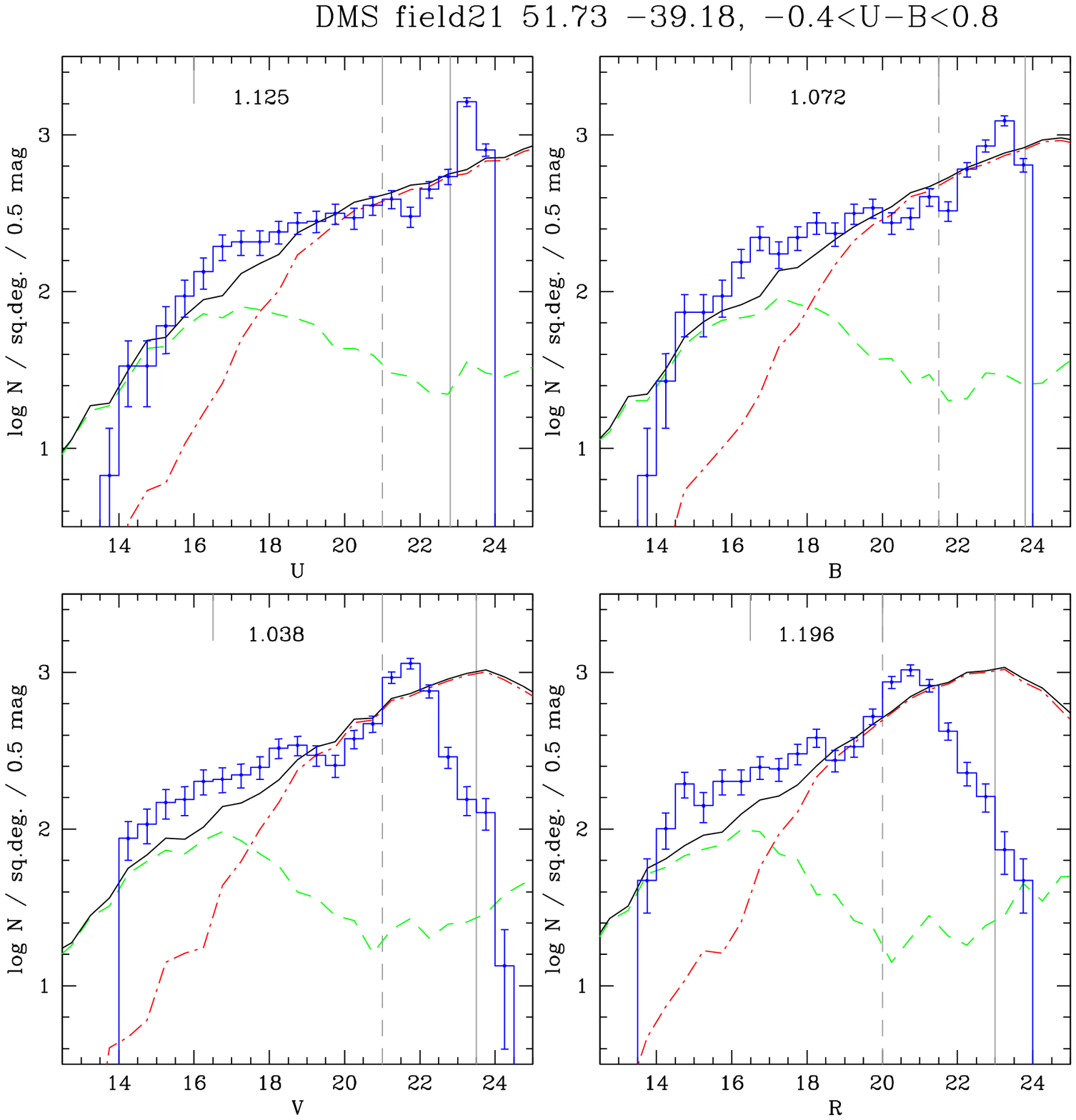}}
	\resizebox{\hsize}{!}{\includegraphics{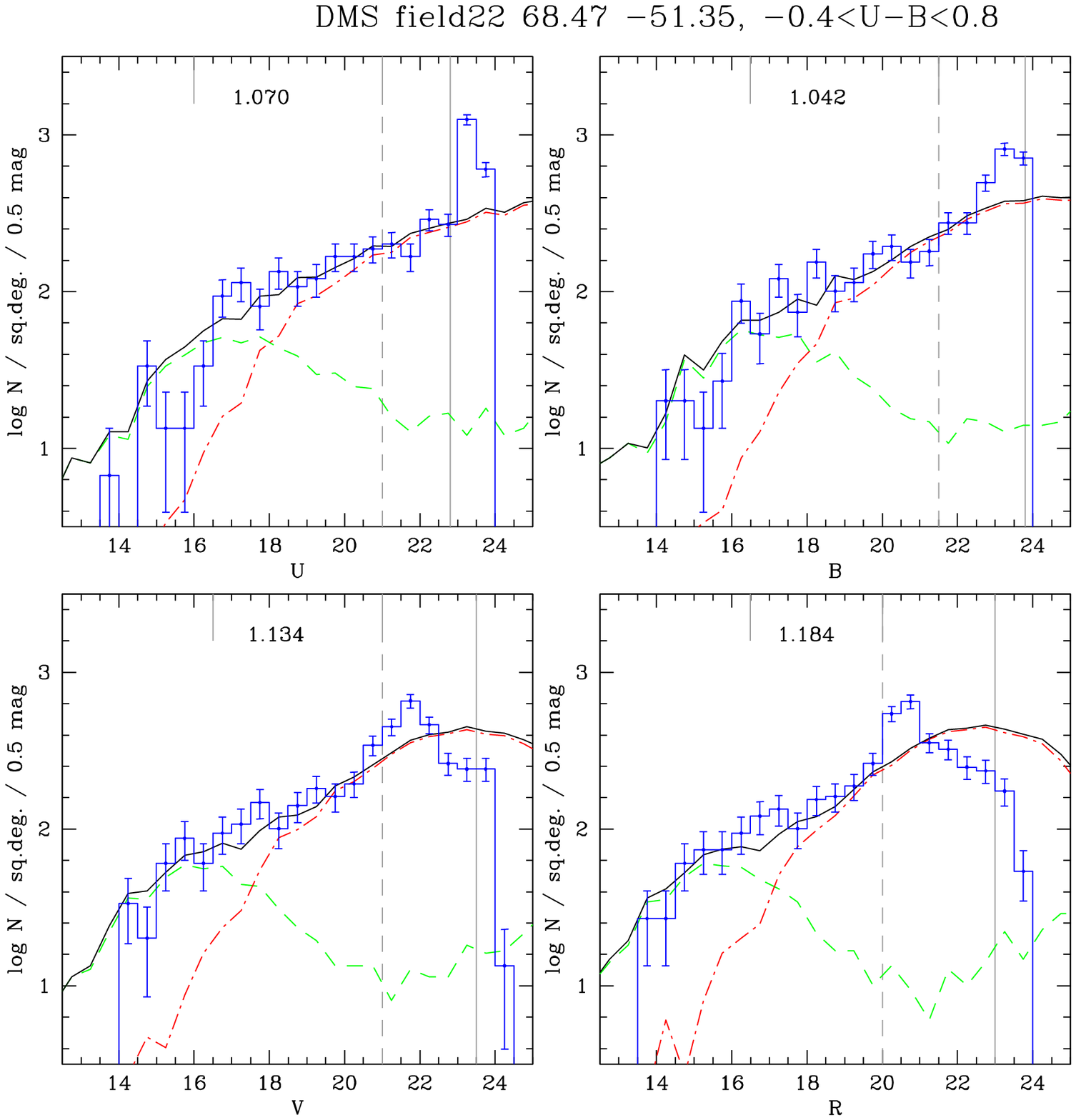}}
	\end{minipage}
	\caption{The same as Fig. \protect\ref{fig_dms},
but now limiting data and models to the ``blue subsample'' with 
$-0.4\le\ub\le0.8$.}
	\label{fig_dms_blue}
	\end{figure*}

Again, the comparison between simulated and
observed number counts for stars in a ``blue subsample'' -- 
now defined in the interval $-0.4\le\ub\le0.8$\footnote{The Johnson
\ub\ colour of DMS separates less the red and blue sequences typical 
of field CMDs, than the \ub\ colour of CDFS -- this latter being based on
ESO WFI filters. This is the reason why we have used a smaller interval of 
\ub\ to define the blue subsample of DMS data, than for CDFS.} 
-- are important for verifying whether we have the right 
proportions between disc and halo stars. A careful examination of 
Fig.~\ref{fig_dms_blue} reveals that this is indeed the case,
but for DMS field 21, where, apparently, simulations
contain too few disc stars. Field 21 is the innermost of DMS fields,
pointing to a Galactic region for which actually our model calibration
indeed presents the largest problems (see Sect.~\ref{sec_2mass}).

\subsubsection{SGP}

\begin{figure}
	\resizebox{\hsize}{!}{\includegraphics{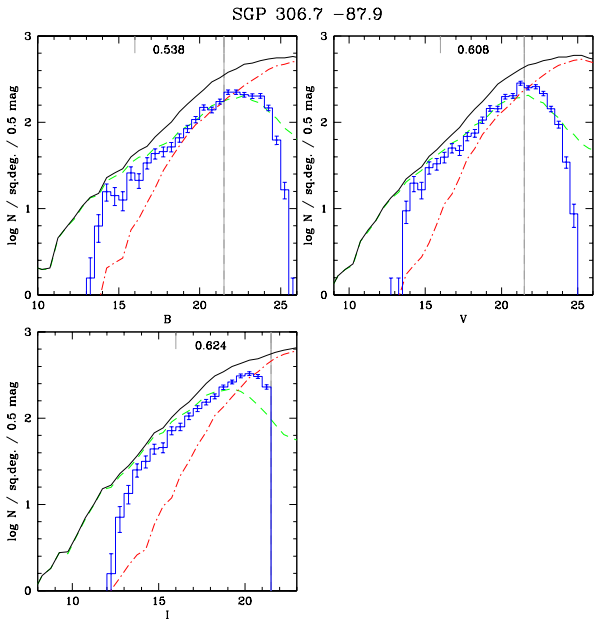}}
	\caption{The same as Fig.~\protect\ref{fig_cdfs},
but for the EIS-deep SGP data (Prandoni et al. 1999). 
The limits of reliability of the data were located,
somewhat arbitrarily this time, at magnitudes 16 and 21.5 for
all filters. }
	\label{fig_sgp}
	\end{figure}

The South Galactic Pole (SGP) as observed by EIS 
(Prandoni et al. 1999; Zaggia et al. 1999) 
presents $BVI$ data for an effective area of 
1.21~deg$^2$ centered at $\ell=306.7\deg, b=-87.9\deg$. 
Fig.~\ref{fig_sgp} shows the comparison between the data 
and our models. This time, the shape of the observed counts are,
in general, fairly well reproduced, except that the amplitude
of our model is twice as large as observed. 
We did not find any obvious way of reducing
this discrepancy, without spoiling the excellent agreement we find
for the other fields considered in this paper -- including 2MASS data
for the SGP itself.
Also, the lack of $U$-band data prevent us from analysing a blue
subsample essential in order to investigate the source 
of the discrepancies.

\subsection{Simulating 2MASS data}
\label{sec_2mass}

As seen above and also in Paper~I, the deep 
simulations of DMS, SGP and CDFS data are ideal for probing
the relative proportions between halo and disc components, as
well as the shape of the halo and the IMF of the disc.
For better probing the disc and its details (spiral arms, dust lanes, 
warps, etc), shallower surveys covering larger areas are better suited. 
However, if one wants to avoid complications
caused by dust, we should consider only counts at the infrared.
In this context, 2MASS constitutes an invaluable dataset: it covers
the all sky in $JH\Ks$ for magnitudes as faint as $J\sim17$.

From the 2MASS All-Sky Data Release Point Source Catalog 
we have selected the sub-sample obeying the so-called 
``2MASS level 1 science requirements'' (see the User's Guide
in Cutri et al.~2003). In practice, these criteria refers to
stellar sources falling outside of tile overlap regions, and 
of high photometric quality (namely ${\rm S/N}>10$ and 
$\sigma<0.11$~mag, band-by-band). For most of the
sky -- excluding the most crowded low-latitude and bulge fields 
-- this subsample of 2MASS is essentially complete for 
magnitudes brighter than about $15$.

\begin{figure*}
	\begin{minipage}{0.48\textwidth}
	\resizebox{\hsize}{!}{\includegraphics{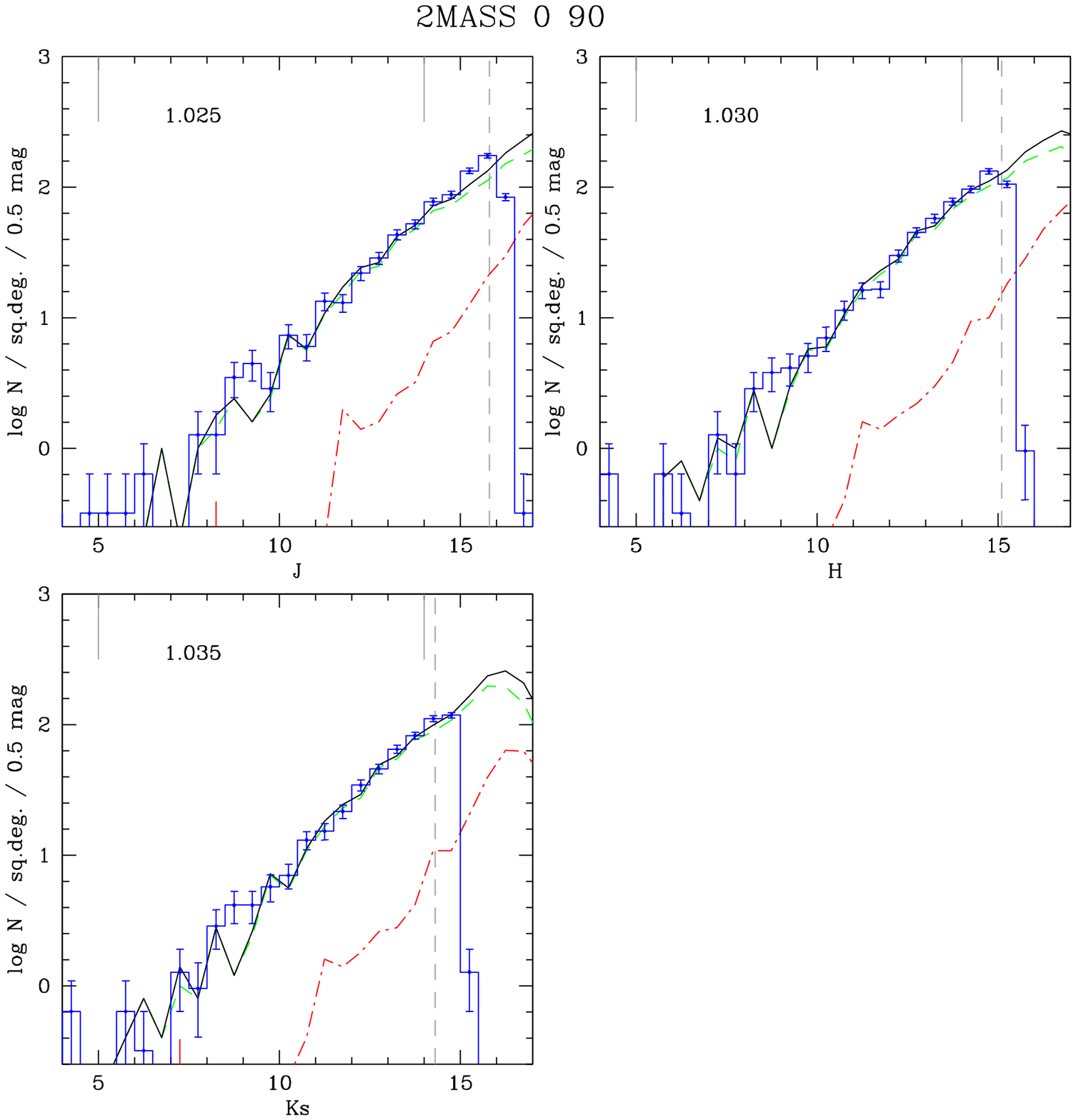}}
	\end{minipage}
	\hfill
	\begin{minipage}{0.48\textwidth}
	\resizebox{\hsize}{!}{\includegraphics{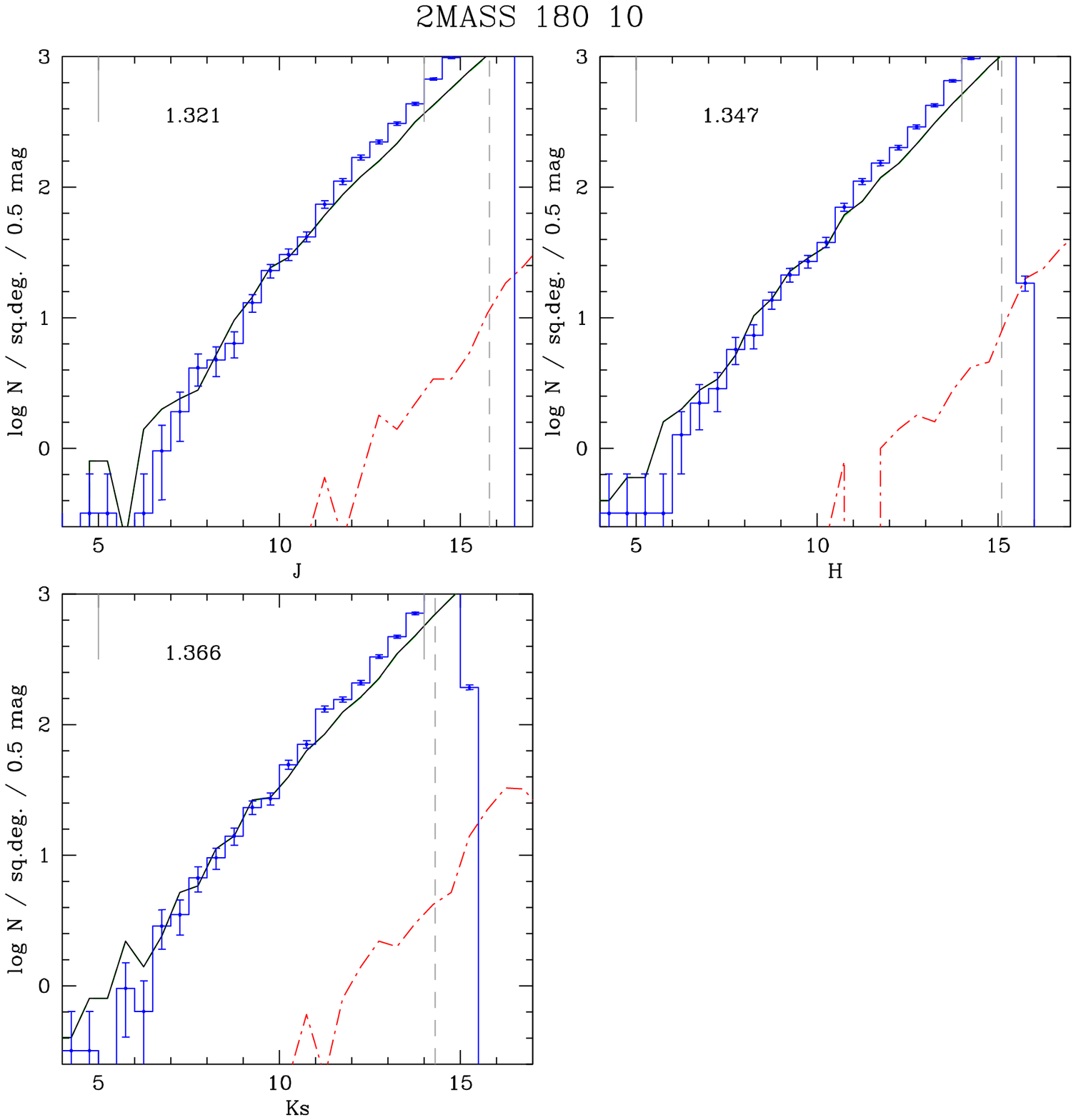}}
	\end{minipage}
	\caption{The same as Fig.~\protect\ref{fig_cdfs},
but for two sample fields of 2MASS data:  
$(\ell=0\deg,b=+90\deg)$ (the NGP), and $(\ell=180\deg,b=+10\deg)$.
Lines have the same meaning as in previous figures.
In both cases, the disc component is responsible for the
bulk of the number counts in the model.}
	\label{fig_2mass_sample}
	\end{figure*}

In the panels of Fig.~\ref{fig_2mass_sample}, we show the 
complete results for two 2MASS fields, one of high latitude
(the NGP at $\ell=0\deg, b=90\deg$) and one of low 
($\ell=180\deg, b=10\deg$). 
Plots for symmetric fields -- namely the SGP 
and the $(\ell=180\deg, b=-10\deg)$ ones -- look very much 
the same and present similar number counts.
The counts in these particular fields are very well
reproduced by the model, over a wide range of magnitudes in all 
the 3 pass-bands of 2MASS. The reader can also notice that the
$J$, $H$ and $\Ks$ diagrams contain essentially the same information, 
so that examining all of them may be redundant. For this reason
and for the sake of conciseness, in the next figure we prefer 
to present just results for the $H$ band.
 
\begin{figure*}
	\begin{minipage}{0.48\textwidth}
	\resizebox{\hsize}{!}{\includegraphics{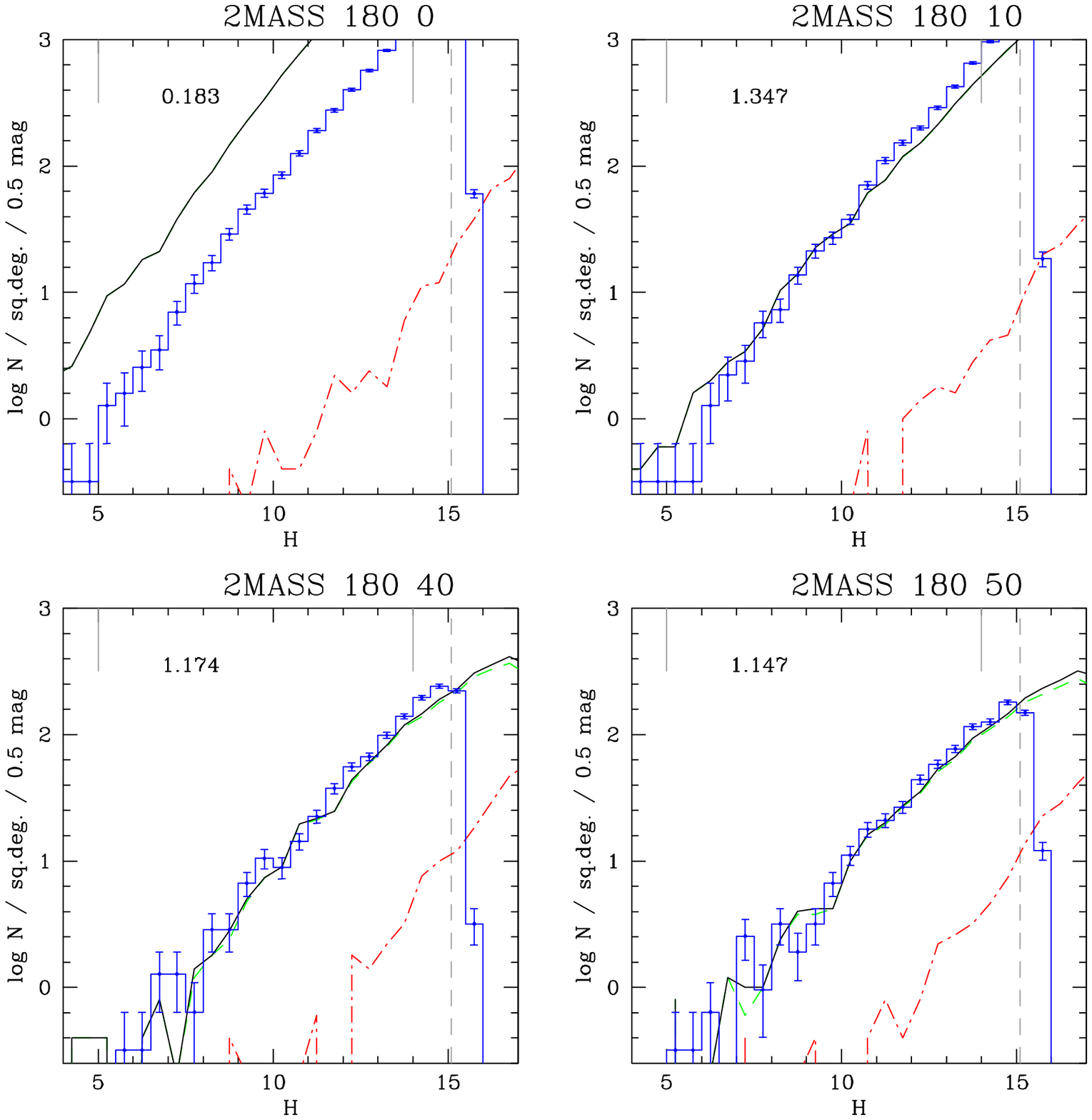}}
	\resizebox{\hsize}{!}{\includegraphics{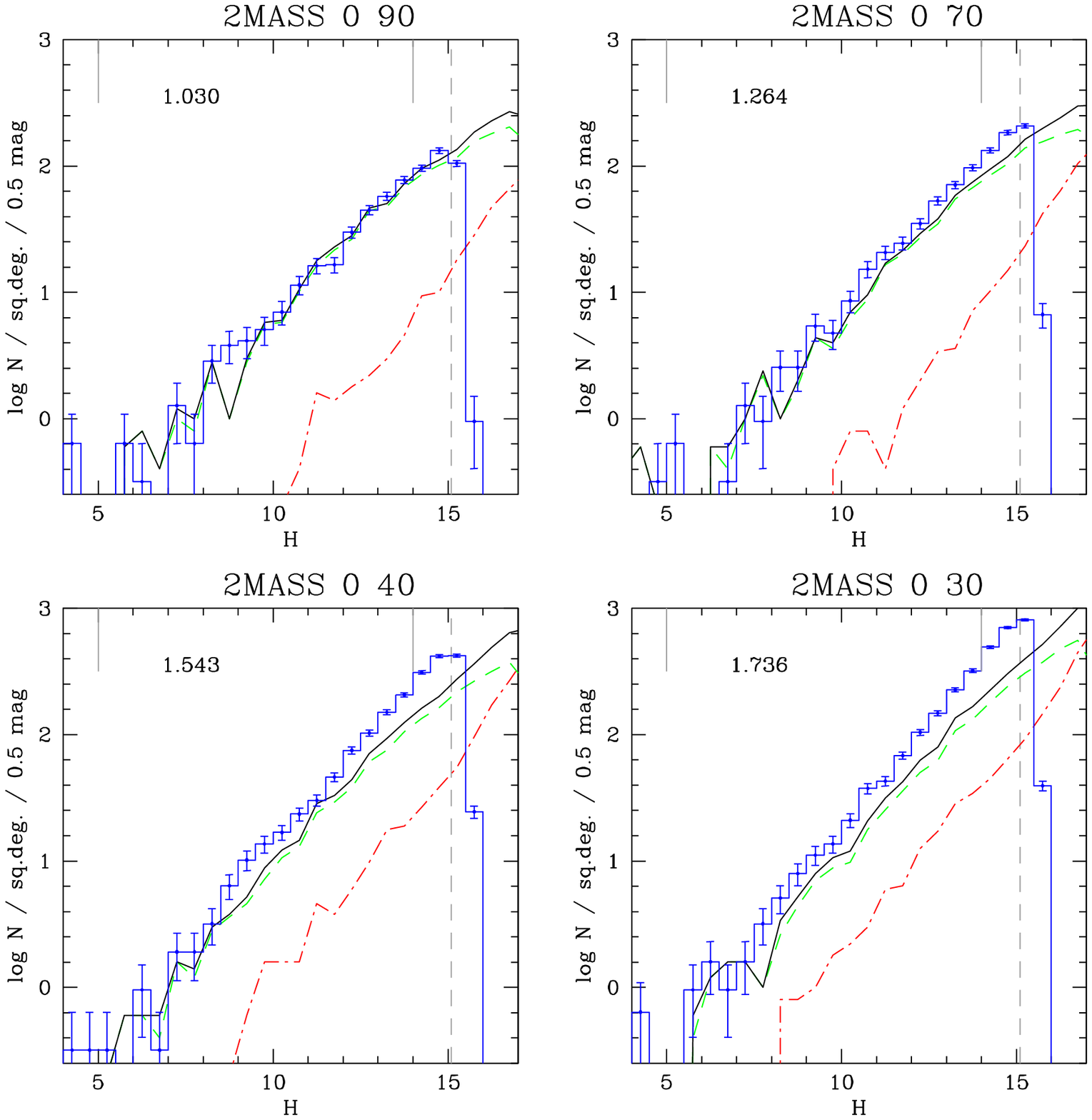}}
	\end{minipage}
	\begin{minipage}{0.48\textwidth}
	\resizebox{\hsize}{!}{\includegraphics{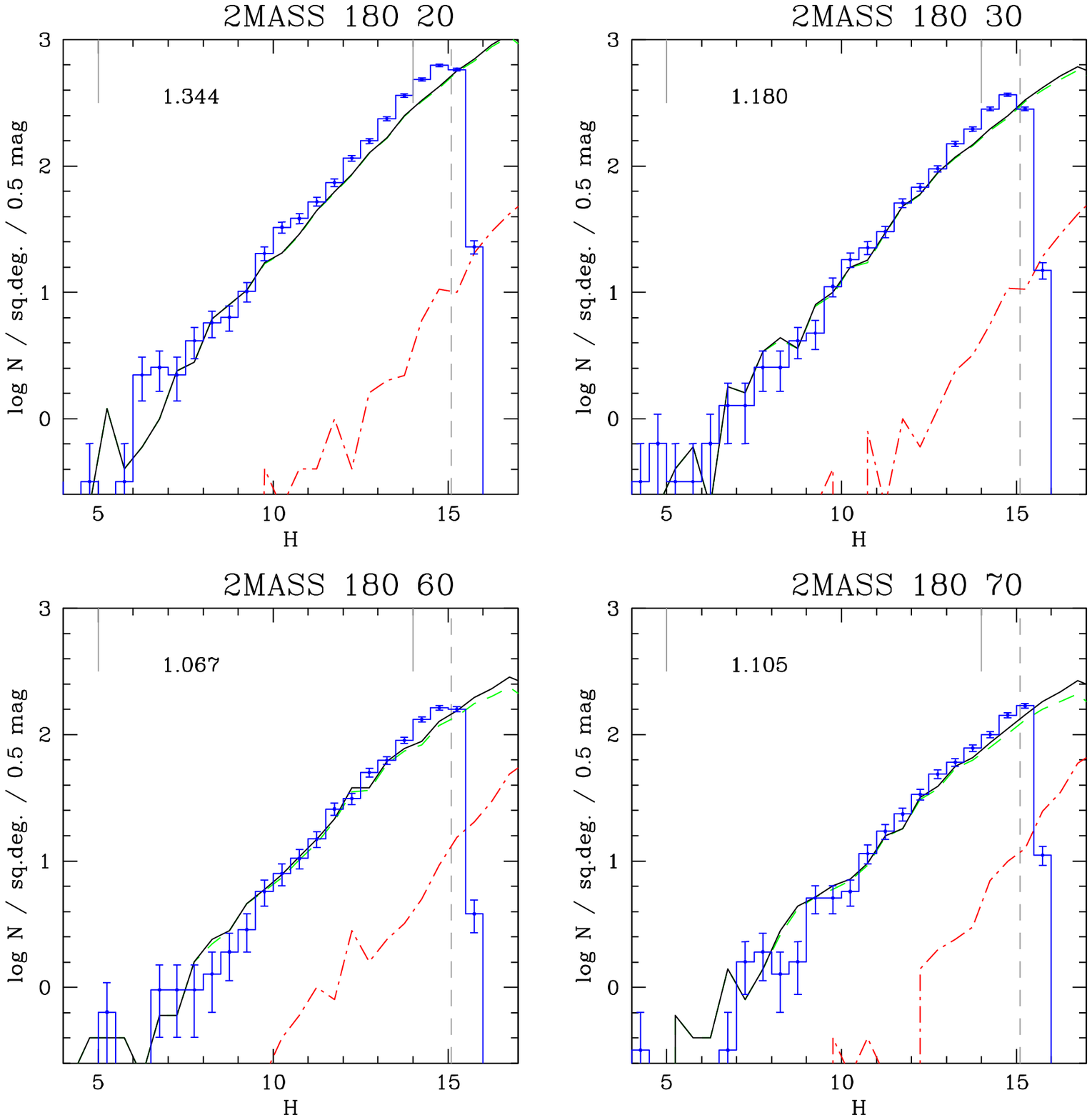}}
	\resizebox{\hsize}{!}{\includegraphics{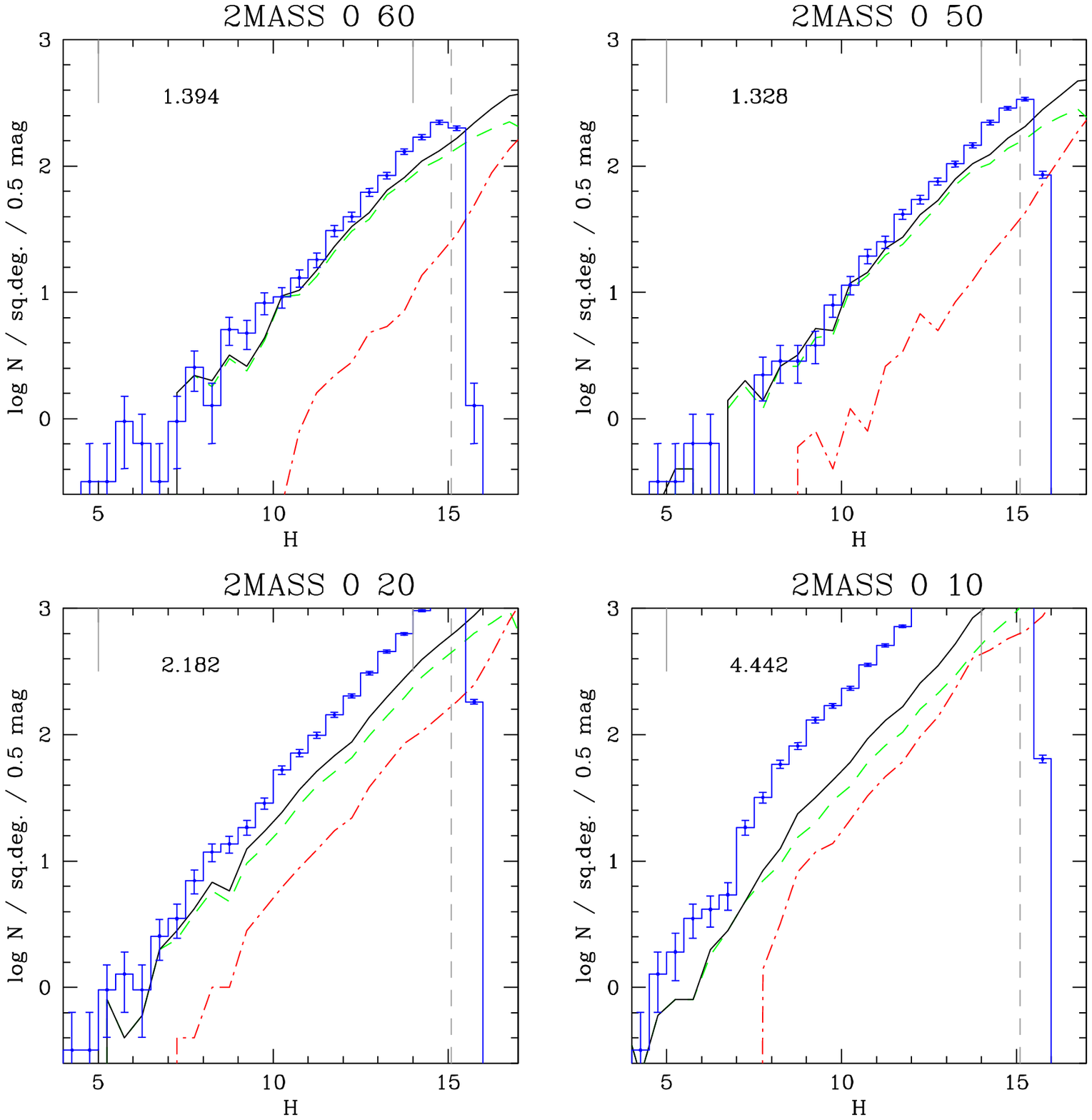}}
	\end{minipage}
	\caption{The same as Fig.~\protect\ref{fig_2mass_sample},
but limited to the $H$-band and for
a series of fields disposed along the  $\ell=0,180\deg$ 
great circle in the sky. Just the northern part of the
circle is presented, since the results are
similar for the southern fields. For the fields 
pointing towards the outskirts of the galactic bulge,
$(\ell=0,b=+20)$ and $(\ell=0,b=+10)$, 
the bulge population is clearly seen as an increase in stellar
counts at $H\ge8$, caused by the bulge RGB, which is not
accounted for by our disc+halo model.}
	\label{fig_2mass_circle}
	\end{figure*}

Fig.~\ref{fig_2mass_circle} presents the $H$-band results
for a series of 2MASS fields disposed along a great circle 
in the sky -- the one at $\ell=0,180\deg$, 
encompassing the Galactic centre, anticentre and 
polar regions. We show only the northern Galaxy fields, since 
results for southern fields are essentially the same. 
\comment{Moreover,
we remind the reader that the 2MASS data for CDFS fields 
have already been presented in previous Fig.~\ref{fig_cdfs}.}

The results are certainly encouraging. We are able to predict the
correct number counts, with errors smaller than $\sim30$ percent, for
all fields located at least 10 degrees above the Galactic Plane,
except for inner Galactic regions where the lack of a Bulge component
-- presently not included in the model -- becomes noticeable.
Moreover, it is not to be excluded that the present description
of the Galactic halo fails for small galactocentric distances,
hence contributing to the discrepancies we find at inner galactic
regions. 

Remarkably, we find that for all 2MASS fields away from the Bulge
we have analysed, disc stars make the bulk of the number counts. 
The halo contribution is almost negligible. 
However, we recall that halo stars contribute to make
a particular feature in wide-area $J$ vs. $J-\Ks$ CMDs, 
namely the central of three vertical fingers discussed by Marigo
et al. (2003), that are clearly present in 2MASS CMDs towards the
LMC. This indicates that the halo component has its importance
in analyses of 2MASS photometry.

In fields too close to the Galactic Plane, instead, 
our model predictions fail, as demonstrated by the
first panel of  Fig.~\ref{fig_2mass_circle} regarding the
direction of the Galactic anticentre. One of the reasons 
for this failure is surely the too simplistic
modelling of the dust distribution along the Galactic Plane.
Improving this aspect of the model, however, is beyond 
the scope of the present paper.

\subsection{Simulating Hipparcos data}
\label{sec_hipparcos}

Examining the stellar counts in the immediate Solar Neighbourhood 
is an obvious test for any Galaxy model. In fact, present models 
usually check whether their results are consistent with the
observed ``local stellar density'', or with some similar parameter  
derived from Hipparcos data (see e.g. Robin et al. 2003).
Of course, by using just a single density parameter as a constraint, 
we ignore the wealth of photometric information that is 
present in the data for nearby stars, that could tell us much about the
distribution of stellar parameters in the disc. In order to start
exploring this information, in the following we try to predict
the counts of a local sample, using the Hipparcos dataset. 

First of all, however, we should remind that 
our model in Paper~I has been effectively calibrated on 
deep data ($V\ga15$).
For such deep surveys the effective distributions of
distance moduli should approach the ones
shown in Fig.~\ref{fig_profile}, i.e. the samples 
are dominated by relatively far objects, at distance moduli 
ranging from $\dmo\sim7$ to $14$ for the disc populations, 
and from $\dmo\sim12$ to $18$ for the halo.
It is then obvious that our previous model
calibration has almost nothing to do with the very 
local sample of stars, i.e.\ the one observed at distances 
lower or comparable to 100~pc, and at bright magnitudes such
as $V\la8$, that we will define below. Deep and local samples
could even be considered, in terms of their 
stellar populations, as completely independent data samples.

\label{sec_sellocalsample}

The Hipparcos and Tycho catalogues (ESA 1997) have provided 
parallaxes with $\sim10$ milliarcsec (mas) accuracy for several 
thousands of stars, together with accurate $BV$ photometry. 
The Hipparcos input catalogue was 
constructed in such a way that there are no clearcut criteria
for defining volume-limited samples out of its data. 
This problem has been recognized by a number of previous authors.
Hernandez et al. (2000) and Bertelli \& Nasi (2001), for instance,
find it to be extremely difficult to 
define volume-limited samples containing enough 
stars for studying the SFR in the solar vicinity up to the
oldest possible ages (see also Schr\"oder \& Pagel 2003). 

However, our aim in this paper is different from previous works.
We consider a subsample of the Hipparcos catalogue to be good provided 
it is complete and could be simulated. Differently from the 
above-mentioned papers, we do not need to limit our simulations 
to stars being all contained in the same volume. 
Our sample can be selected by using a few simple criteria, 
based on the following realizations: 
\bite
\item the Hipparcos catalogue contains all stars brighter than 
$V_{\rm lim}\simeq7$;
\item from them, all parallaxes $\pi$ higher 
than $\pi_{\rm lim}\simeq10$ mas have been measured; 
\item due to the presence of parallax errors, the sample
defined by $\pi_{\rm lim}$ comprehends a volume of radius
$r_{\rm lim}$ somewhat larger than $1/\pi_{\rm lim}\simeq100$ pc;
\item among these stars, the bulk of binaries has been 
identified.
\eite
Therefore, we select from the Hipparcos catalogue all stars 
with $V<V_{\rm lim}$ {\em and} $\pi>\pi_{\rm lim}$, and not 
classified as binaries of any kind. $V_{\rm lim}$ and 
$\pi_{\rm lim}$ are kept as parameters.
Initially, we have conservatively adopted 
$V_{\rm lim}=7$ and $\pi_{\rm lim}=10$ mas, 
which corresponds to $r_{\rm lim}>100$ pc.

\label{sec_simlocalsample}

We then simulated this local sample using the TRILEGAL code.
To do so, we have generated synthetic samples up to a distance
$r_{\rm lim}=200$ pc. This is large enough to include all stars that,
due to parallax errors, will later be scattered to apparent distances
closer than 100~pc.

The simulated true physical distances $r_0$ are first converted
in the true parallax $\pi_0=1/r_0$. 
The simulated parallax errors $\delta\pi$ (described in the 
Appendix) are then added to $\pi_0$ so as to
generate the ``observed'' parallaxes and distances, $\pi$ and 
$r=1/\pi$. The ``observed'' absolute magnitude is then derived
by the usual formula, $M_V=V-5\log r +5$.
Extinction has been ignored, since its effect 
inside a radius of 200 pc is negligible.

	\begin{figure*}
	\resizebox{0.8\hsize}{!}{\includegraphics{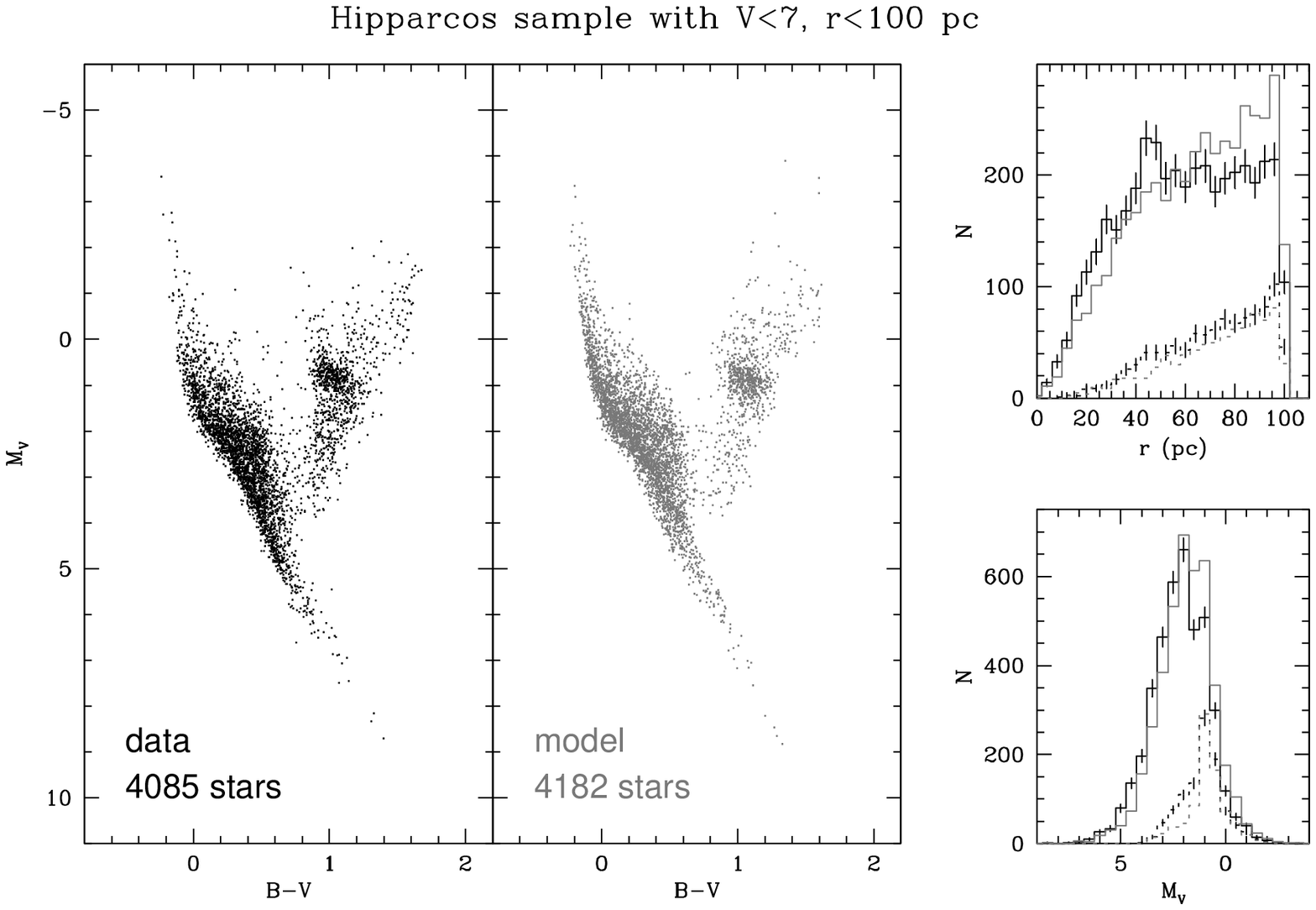}}
	\caption{The several panels show our Hipparcos simulation 
(grey) versus the real data (dark). On the {\bf left panels}, we have the
\mv\ vs.\ \bv\ diagram of both samples (observed and simulated) 
limited to a parallax of $\pi<10$~mas (apparent distance $r=1/\pi<100$~pc)
and an apparent magnitude of $V<7$. The agreement between simulated and 
observed samples is remarkable for most regions of the diagram.
The lines in the {\bf right panels} show the corresponding
distributions of apparent distance $r$
({\bf top}) and derived absolute magnitude ({\bf bottom}),
for both the total samples (continuous lines) and the 
subsample of subgiants and giants (dotted lines). 
For the data histograms, error bars denote the standard 
error ($\sqrt{N}$) of a Poisson distribution, 
and serve as a guide to the comparison between model and data. 
As can be noticed, the simulation predicts about the same total 
star counts as observed, with just a 2.5 percent excess of 
model stars. Looking at the right panels, however, we notice
a significant excess of simulated stars at 
$r\simeq80$~pc, a deficiency at $r\simeq20$~pc, 
and the obvious failure to reproduce the spike at 
$r=45$~pc that corresponds to the Hyades cluster. 
Moreover, in the comparison of number counts as a function of \mv,
there seems to be an excess of bright stars and a deficit of
faint ones.}
	\label{fig_panelhipp7}
	\end{figure*}

	\begin{figure*}
	\resizebox{0.8\hsize}{!}{\includegraphics{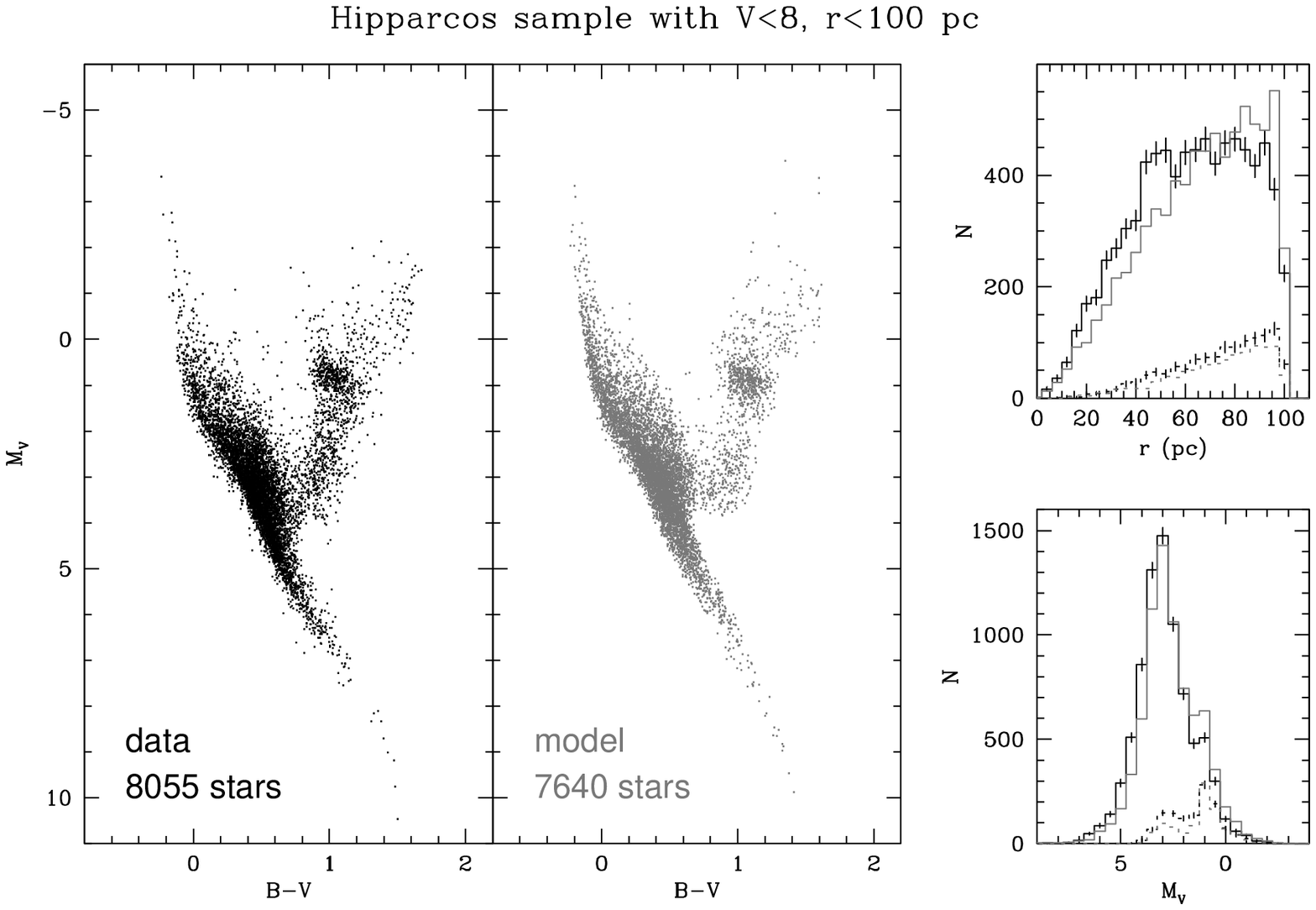}}
	\caption{The same as Fig.~\protect\ref{fig_panelhipp7},
but for samples limited to an apparent magnitude of $V_{\rm lim}<8$.
Notice the increased number counts of stars with $\mv>2$.}
	\label{fig_panelhipp8}
	\end{figure*}

\label{sec_reslocalsample}

The results of this exercise can be seen in 
Fig.~\ref{fig_panelhipp7}, for the conservative 
choice of $V_{\rm lim}=7$ and $\pi_{\rm lim}=10$ mas.
In the left-hand panel, we compare the simulated and observed
\mv\ vs. \bv\ diagram.  
The agreement between simulated and observed samples is striking.
It can be noticed that models describe very well both the location
and width of the main sequence all along from $\mv\sim9$ to $-4$. 
Particularly good is the description of the left boundary of the
MS, very well marked because it is produced by stars in their phase of
slowest evolution, when they depart from their ZAMS to the right
in the HR diagram. Regarding the MS width,
we know from stellar models that it is affected essentially 
by two factors: the assumed metallicity dispersion (or equivalently 
the AMR in our models) and the efficiency of convective core 
overshooting for $M>1$~\Msun\ stars. The good agreement 
between models and simulations seems to indicate that both these
ingredients are well described in our models. Of course,  
before considering the implications that this result may have
for the disc AMR and for the theory of stellar evolution more
quantitative comparisons would have to be made. However, such a
discussion is beyond the scope of the present paper.

Regarding the evolved stars (subgiants and red giants), the agreement
is also very good: we can notice the right width of the subgiant 
and lowest part of RGB; the clumping of core-He burning stars in the
right location; the bifurcation of the red giants above the clump
into two loose broad sequences: a vertical one made of intermediate-mass 
core-He burners and the inclined one, going to the red, made of 
genuine first-ascent RGB and early-AGB stars. 

As can be seen, the simulation predicts about the same star 
counts as observed: the total number of objects in both panels 
is 4085 and 4182 for the data and models, respectively.

The right panels show the corresponding
distributions of apparent distance $r$ and derived absolute 
magnitude $M_V$ (continuous lines). The dashed lines 
refer to the subsample of subgiants and giants, defined by
the stars with $\mv>6.82\times(\bv)-2$. 

There, although a first look indicates a good overall level
of agreement between models and data, some discrepancies 
become apparent. In the histogram of $r$, the most noticeable
one seems to be a modest excess of simulated stars at 
$r\sim80$~pc, that amounts to about 20 percent. 
Since the volume sampled in the simulation is very small,
we consider that such a discrepancy is unlikely to be derived 
from inhomogeneities in the local distribution of stars;
more likely, a better simulation of Hipparcos parallax errors 
could improve the models in this particular range 
of distances (parallaxes). There is also a modest
deficit of simulated stars in the smallest distances 
(from 15 to 35 pc), amounting again to about 20 percent, 
which however does not appear among the subgiants and giants. 
This again might indicate a problem in the simulation of
parallax errors for the faintest stars. On the other hand, 
the spike of observed star counts at $r=45$~pc is to be 
assigned to the Hyades cluster, which was not considered in 
our simulation. 

The histogram of number counts against \mv\ indicates, once again, 
some modest discrepancies, that are however statistically 
very significant. The most important one consists on an excess of 
simulated bright stars accompanied by a deficit of the 
faintest ones. We have performed a KS-test comparison between
the two \mv\ distributions, and find that the probability of 
them being drawed from the same distribution would be highly 
increased -- from its present close-to-zero value up to about 0.3 --
if our models were shifted by 0.26~mag in \mv. We think however
that applying such a shift whould not be justified, and it could 
not be easily translated onto a physical interpretation 
(i.e. shifts in the \mv\ distribution can be forced by using 
models with a corrected photometric zero-point, different IMF, 
different SFR, modified prescription for simulating parallax 
errors, or a combination of all these effects). 

To conclude, we remark that all the above-mentioned 
discrepancies are, from the statistical point of view, 
highly significant, since they refer to samples
containing large numbers of stars. They may be indicating 
points where the present models can be improved.
They however do not invalidate the present model calibration,
for a series of reasons: first of all, many of the discrepancies 
are suspected to result from the imperfect simulation of
parallax errors; second, when one deals with normal
star counts data, fine details (i.e. those seen at $\sim0.25$ mag
resolution) of the \mv\ distribution become of little relevance 
since stars are dispersed over a large range of 
distance moduli; third, the most relevant comparison regards 
the total star counts that are directly linked to the local 
density of stars: in our case they are very well predicted, 
to within 5 percent. Reaching such a result for the star counts
is already remarkable, whereas reaching a statistically-robust
comparison with Hipparcos, in all its details, may be still 
far from reach. In this regard, the present work represents 
just the first attemp.

Figure~\ref{fig_panelhipp8} presents the same as 
Fig.~\ref{fig_panelhipp7}, but for a slightly deeper sample, 
of $V_{\rm lim}=8$ and $\pi_{\rm lim}=10$ mas.
The effect of selecting a deeper $V_{\rm lim}$ is that more stars
with $\mv>2$ are included in the sample, then increasing the
contribution of intermediate-age to old stars (in their main sequence,
subgiant branch, and initial RGB evolution) to the number counts.
In this case, the star counts are 8055 and 7640 
for data and models, respectively. The discrepancies between
model and observations seem slightly increased, as expected 
since we are including data for which the completeness starts 
to become an issue, and for which parallax errors are slightly
larger with respect to the previous $V<7$ sample.

\section{Discussion and conclusions}
\label{sec_conclu}

We have presented a new code for simulating the photometry of 
Galaxy fields. 

The code has been calibrated by predicting counts in a variety 
of stellar surveys, that comprehend some very deep multi-passband 
catalogues cleaned from galaxies (CDFS, DMS, SGP), 
the ``intermediate-depth'' near-IR point source catalogue of 2MASS, 
and the very local stellar sample derived from Hipparcos catalogue. 

The results are certainly satisfactory, since we have demonstrated
that the predicted star counts agree well with the observed ones. 
The typical discrepancies are smaller than 30 percent for most of 
the sky, and inside the estimated magnitude limits of reliability 
of the observed star counts.
This agreement is remarkable, when we consider the wide ranges of
magnitudes, passbands (from $U$ to $K$) and sky positions that were 
considered in this work.

The major discrepancies were found for: (1) Inner Galactic fields,
located within about 30 degrees from the Galactic Center, for which we
largely underestimate the number counts. Part of this discrepancy 
can be attributed to the lack of a bulge component in the 
present model, but probably a better modelling of the inner disc
and halo densities is also necessary. (2) Low-latitude fields,
with $|b|\la10\deg$, which probably require a detailed modelling of
the distribution of dust and recent star formation along the disc. 
Finally, (3) the SGP as observed by EIS, for which we predict 
twice as many counts in the optical as observed. In this case,
the origin of the discrepancy could not be identified.

Note that the present model calibration is not yet fully optimized 
and is likely not to be unique, in the
sense that other choices for the stellar densities and 
star formation histories of Galaxy components might produce 
similarly good results. In fact, the question arises whether 
TRILEGAL could be adapted to find, in an objective way, 
a maximum-likelihood solution for 
the functions $\rho$, $\psi$, $\phi$, etc. -- 
using a database of few high-quality 
multiband catalogues covering several regions of the sky and 
with a large range in depth. The answer is likely
yes, but the way is certainly not straightforward. 
Such a work would imply at least the following steps:
(a) finding a zeroth-order calibration producing 
``acceptable'' results, which is actually the step performed 
in this paper;
(b) establishing a robust likelihood criterium for 
comparing the resulting models with the stellar data;
(c) establishing a numerical algorithm to migrate from the 
zeroth-order to improved maximum-likelihood solutions
(see for instance Ng et al. 2002);
(d) exploring the problem of uniqueness of solution
by using different starting solutions.
Therefore, what is presented in this paper can be seen as the
initial step of a bigger project, that we expect to pursue
in the future.

Aside from the good model calibration we have reached,
the most important advantages of the TRILEGAL code can be identified 
in: (1) the
fairly complete database of stellar evolutionary tracks already
implemented; (2) the use of an extended spectral library to 
simulate many different photometric systems, and their extinction
coefficients, in a self-consistent way,
(3) the modular and flexible structure of the code, that allows 
easy changes and additions to both input functions (SFR, AMR, IMF,
etc) and geometric parameters (the density of Galaxy components,
Sun's position, pointing parameters, etc.).

With respect to other population synthesis codes commonly used to 
simulate the photometry of Galaxy fields (e.g. Ng et al. 1995, 
Vallenari et al. 2000, Castellani et al. 2002, Robin et al. 2003), 
TRILEGAL shares the advantage of being intrinsically self-consistent
in the relative numbers of stars predicted to be in different 
evolutionary phases (including stellar remnants such as 
white dwarfs). In fact, for a given Galaxy geometry, stellar 
number ratios are univoquely determined by the choice of SFR, 
AMR and IMF, and are not tunable parameters.
In TRILEGAL, this self-consistency of population synthesis codes
is kept as a 
very stringent criteria, since there are explicit checks for the
continuity of all stellar quantities (including 
core mass, envelope mass, and surface chemical composition when 
applicable) in the isochrone-construction routines that make part 
of the code.

As already mentioned, a 
main positive characteristic of TRILEGAL consists 
in the extreme flexibility in the way input libraries 
(evolutionary tracks, atmospheres)
and functions (geometry, IMF, SFR, AMR of Galaxy components) 
can be changed, tested,
and added to a database for future use. Improvements in the
stellar evolutionary tracks, for instance those described in 
Marigo et al.~(2003), have been inserted in TRILEGAL
in almost no time. We have so far computed test models 
in at least 10 different photometric systems (including
$UBVRIJHK$, Washington, HST-based instruments like WFPC2, NICMOS 
and ACS, the EIS photometric system, 2MASS and SDSS), and we are 
confident that virtually any broadband Vega, AB or ST magnitude
system can be considered as well (cf. Girardi et al.~2002). 
Needless to say, 
before the present TRILEGAL calibration has been considered as 
acceptable, we have made wide use of its flexibility by testing
many different IMFs, AMRs, SFRs, and density 
functions published by different authors.
Such a flexibility is of fundamental importance for facing 
the huge amount of wide-field photometric data that is becoming  
available these days, and to take immediate advantage of
the next generation of improved/expanded stellar evolutionary 
and atmospheric models, that are now being prepared by different
groups around the world. 

Therefore, the TRILEGAL code is ready to use for the variety of 
wide-angle surveys in the optical/infrared that will become available 
in the coming years and will provide constraints that will help us to 
pin down the structure of our Galaxy.

\begin{acknowledgements}
This publication makes use of data products from the 
Two Micron All Sky Survey, which is a joint project of the 
University of Massachusetts and the Infrared Processing and 
Analysis Center/California Institute of Technology, 
funded by the National Aeronautics and Space Administration 
and the National Science Foundation.
We acknowledge the referee for the interesting remarks,
which helped to improve this paper.
\end{acknowledgements}


\section*{References}

\begin{description}
\bibitem[]{}  Allard, F., Hauschildt, P.H., Alexander, D.R., Tamanai, A., 
	Ferguson, J.W. 2000, in ASP Conf.\ Series v. 212, 
	From giant planets to cool stars, eds. C.A. Griffith \& M.S. Marley,
	127 
\bibitem[]{} Arnouts, S., Vandame, B., Benoist, C., et al. 2001, A\&A, 379, 740
\bibitem[]{} Bahcall, J.N. 1986, ARA\&A, 24, 577
\bibitem[]{} Bahcall, J.N., \& Soneira, R.M., 1980, ApJS, 44, 73
\bibitem[]{} Bahcall, J.N., \& Soneira, R.M., 1984, ApJS, 55, 67
\bibitem[]{} Barmina, R., Girardi, L., Chiosi, C. 2002, A\&A, 385, 847 
\bibitem[]{} Benvenuto, O.G., \& Althaus, L.G. 1999, MNRAS, 303, 30
\bibitem[]{} Bertelli, G., \& Nasi, E.\ 2001, AJ, 121, 1013 
\bibitem[]{} Bertelli, G., Bressan, A., Chiosi, C., Fagotto, F., Nasi, E. 
	1994, A\&AS 106, 275
\bibitem[]{} Bessell, M.S., Castelli, F., Plez, B. 1998, A\&A, 333, 231
\bibitem[]{} Binney, J., Gerhard, O., Spergel, D. 1997, MNRAS, 288, 365
\bibitem[]{} Cardelli, J.A., Clayton, G.C., Mathis, J.S. 1989, ApJ, 345, 245
\bibitem[]{} Carraro, G., Girardi, L., Marigo, P. 2002, MNRAS, 332, 705
\bibitem[]{} Castellani, V., Cignoni, M., Degl'Innocenti, S., Petroni, S., 
	Prada Moroni, P.G. 2002, MNRAS, 334, 69
\bibitem[]{} Castelli, F., Gratton, R.G., Kurucz, R.L. 1997, A\&A, 318, 841
\bibitem[]{} Chabrier, G. 2001, ApJ, 554, 1274
\bibitem[]{} Chabrier, G., Baraffe, I., Allard, F., Hauschildt, P.H. 2000,
	ApJ, 542, 464
\bibitem[]{} Chen B., Stoughton C., Smith J.A., et al. 2001, ApJ, 553, 184
\bibitem[]{} Cohen, M. 1995, ApJ, 444, 874
\bibitem[]{} Cutri, R.M., Skrutskie, M.F., Van Dyk, S., et al.
	2003, All-Sky Data Release Explanatory Supplement,
	{\tt http://www.ipac.caltech.edu/\-2mass/\-releases/\-allsky/\-doc/\-explsup.html}
\bibitem[]{} de Vaucoulers, G. 1959, in Handbuch der Physik, ed. S. Flugge,
	Springer-Verlag, Berlin, v. 53, p.\ 311.
\bibitem[]{} ESA, 1997, The Hipparcos and Tycho Catalogues
\bibitem[]{} Finley, D.S., Koester, D., Basri, G. 1997, ApJ, 488, 375
\bibitem[]{} Fluks, M.A, Plez, B., The, P.S., et al. 1994, A\&AS, 105, 311
\bibitem[]{} Fuhrmann, K. 1998, A\&A, 338, 161
\bibitem[]{} Gilmore, G. 1984, MNRAS, 207, 223
\bibitem[]{} Gilmore, G., \& Reid, N. 1983, MNRAS, 202, 1025
\bibitem[]{} Girardi, L. 1997, PdD Thesis, Universidade Federal do 
	Rio Grande do Sul, Porto Alegre, Brazil
\bibitem[]{} Girardi, L. 1999, MNRAS, 308, 818
\bibitem[]{} Girardi, L., \& Bertelli, G. 1998, MNRAS, 300, 533
\bibitem[]{} Girardi, L., \& Salaris, M., 2001, MNRAS, 323, 109
\bibitem[]{} Girardi, L., Bressan, A., Chiosi, C., Bertelli, G., Nasi, E. 1996,
	A\&AS, 117, 113
\bibitem[]{} Girardi, L., Groenewegen, M.A.T., Weiss, A., Salaris, M. 
	1998, MNRAS, 301, 149
\bibitem[]{} Girardi, L., Bressan, A., Bertelli, G., Chiosi, C. 2000, 
	A\&AS, 141, 371
\bibitem[]{} Girardi, L., Bertelli, G., Bressan, A., et al. 2002, 
	A\&A, 391, 195 
\bibitem[]{} Girardi, L., Bertelli, G., Chiosi, C., Marigo, P. 2003, 
	IAUS, 212, 551 
\bibitem[]{} Girardi, L., Grebel, E.K., Odenkirchen, M., Chiosi, C. 2004,
        A\&A, 422, 205
\bibitem[]{} Grebel, E.K., \& Roberts, W.J. 1995, A\&AS, 109, 293
\bibitem[]{} Groenewegen M.A.T., Girardi L., Hatziminaoglou E., et al. 2002, 
	A\&A, 392, 741 (Paper~I)
\bibitem[]{} Hall, P.B., Osmer, P.S., Green, R.F., Porter, A.C., Warren, S.J.
	1996, ApJS, 104, 185
\bibitem[]{} Hatziminaoglou, E., Groenewegen, M.A.T., da Costa, L., 
	et al. 2002, A\&A, 384, 81 
\bibitem[]{} Haywood, M., 1994, A\&A, 282, 444
\bibitem[]{} Henry, R.B.C. \& Worthey G., 1999, PASP, 111, 919
\bibitem[]{} Hernandez, X., Valls-Gabaud, D., Gilmore, G. 2000, 
	MNRAS, 316, 605
\bibitem[]{} Holmberg, J., Flynn, C. 2004, MNRAS, 352, 440
\bibitem[]{} Homeier, D., Koester, D., Hagen, H.-J., et al. 1998, 
	A\&A, 338, 563
\bibitem[]{} Kroupa, P. 2001, MNRAS, 322, 231
\bibitem[]{} Kurucz, R.L. 1993, in IAU
     	Symp. 149, The Stellar Populations of Galaxies, 
	eds.\ B.\ Barbuy, A.\ Renzini, Dordrecht, Kluwer, 225
\bibitem[]{} Larson, R.B. 1986, MNRAS, 218, 409
\bibitem[]{} Lyng\aa, G. 1982, A\&A, 109, 213
\bibitem[]{} Marigo, P. 2002, A\&A, 387, 507
\bibitem[]{} Marigo, P., Girardi, L., Chiosi, C. 2003, A\&A, 403, 225
\bibitem[]{} Ma\'\i z-Apell\'aniz, J. 2001, AJ, 121, 2737
\bibitem[]{} M\'endez, R.A., van Altena, W.F. 1996, AJ, 112, 655
\bibitem[]{} M\'endez, R.A., van Altena, W.F. 1998, A\&A, 330, 910
\bibitem[]{} Ng, Y.K., Bertelli, G., Bressan, A., Chiosi, C., Lub, J. 1995, 
	A\&A, 295, 655
\bibitem[]{} Ng, Y.K., Bertelli, G., Chiosi, C., Bressan, A. 1997, 
	A\&A, 324, 65
\bibitem[]{} Ng, Y.K., Brogt, E.,  Chiosi, C., Bertelli, G. 2002, 
	A\&A, 392, 1129
\bibitem[]{} Ojha, D.K. 2001, MNRAS, 322, 426
\bibitem[]{} Osmer, P.S., Kennefick, J.D., Hall, P.B., Green, R.F. 1998, 
	ApJS, 119, 189
\bibitem[]{} Parenago, P.P. 1945, Russian AJ, 22, 129
\bibitem[]{} Perryman, M.A.C., et al., 1997, A\&A, 323, L49
\bibitem[]{} Prandoni, I., Wichmann, R., da Costa L., et al. 
	1999, A\&A, 345, 448
\bibitem[]{} Rana, N.C., \& Basu, S. 1992, A\&A, 265 ,499
\bibitem[]{} Reid, N., \& Majewski, S.R. 1993, ApJ, 409, 635
\bibitem[]{} Renzini, A., da Costa, L.N. 1997, The Messenger, 87, 23
\bibitem[]{} Robin, A.C., \& Cr\'ez\'e, M. 1986, A\&A, 157, 71
\bibitem[]{} Robin, A.C., Reyl\'e, C., Cr\'ez\'e, M. 2000, A\&A, 359, 103
\bibitem[]{} Robin, A.C., Reyl\'e, C., Derri\'ere, S., Picaud, S. 2003, 
	A\&A, 409, 523
\bibitem[]{} Rocha-Pinto, H.J., Maciel, W.J., Scalo, J., Flynn, C. 2000, 
	A\&A, 358, 850
\bibitem[]{} Ryan, S.G., \& Norris, J.E. 1991, AJ, 101, 1865
\bibitem[]{} Salaris, M., \& Weiss, A. 1998, A\&A, 335, 943
\bibitem[]{} Salaris, M., Chieffi, A., Straniero, O. 1993, ApJ, 414, 580
\bibitem[]{} Salasnich, B., Girardi, L., Weiss, A., Chiosi, C. 2000, 
	A\&A, 361, 1023
\bibitem[]{} Salpeter, E.E. 1955, ApJ, 121, 161
\bibitem[]{} Schlegel, D.J., Finkbeiner, D.P., Davis, M. 1998, ApJ, 500, 525
\bibitem[]{} Schr\"oder, K.-P., Pagel, B.E.J. 2003, MNRAS 343, 1231
\bibitem[]{} Vallenari, A., Bertelli, G., Schmidtobreick, L. 2000, A\&A, 
	364, 925
\bibitem[]{} Vandame, B., Olsen, L.F., Jorgensen, H.E., et al. 
	2001, astro-ph/0102370 
\bibitem[]{} VandenBerg, D.A. 2000, ApJS, 129, 315
\bibitem[]{} Vassiliadis, E., Wood, P.R. 1994, ApJS, 92, 125
\bibitem[]{} York, D.G., et al., (the SDSS collaboration), 2000, AJ, 120, 1579
\bibitem[]{} Young, P.J. 1986, AJ, 81, 807
\bibitem[]{} Zaggia, S., Hook, I., Mendez, R., et al. 1999, A\&AS, 137, 75
\bibitem[]{} Zheng, Z., Flynn, C., Gould, A., Bahcall, J.N., Salim, S. 2001, 
	ApJ, 555, 393
\end{description}  

\appendix

\section{Simulation of parallax errors in Hipparcos}

\begin{figure}
	\resizebox{\hsize}{!}{\includegraphics{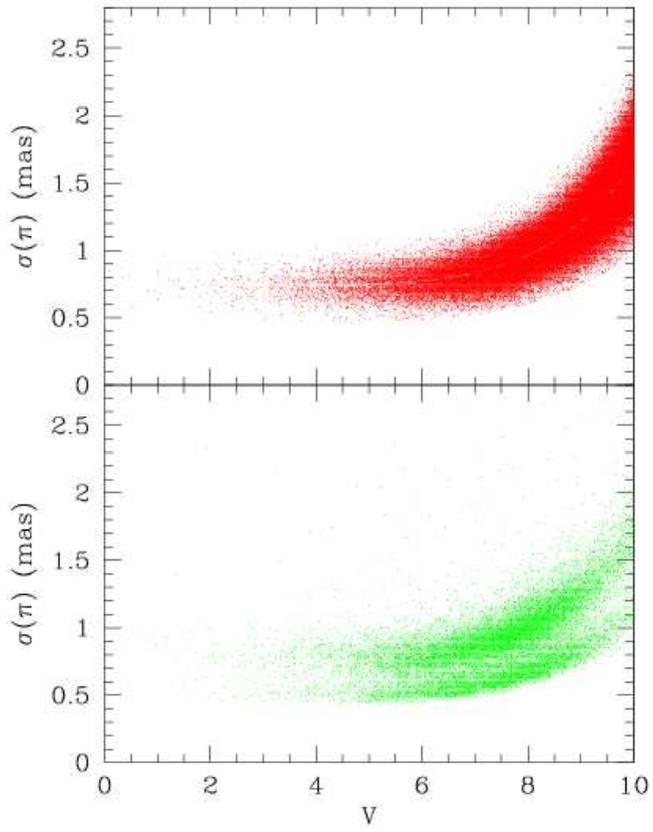}}
	\caption{Errors in Hipparcos parallaxes. The data is 
in the lower panel, our simulations in the upper one.}
	\label{fig_parerr}
	\end{figure}

In this Appendix we describe how the parallax errors in the Hipparcos
catalogue have been simulated. From the about 118000 objects in the
Hipparcos catalogue the about 99000 objects have been selected that
fulfill: ``goodness-of-fit'' flag (H30) less than 3,
``percentage-of-rejected-data'' flag (H29) less than 10, ``number of
components'' flag (H58) of 1, that have a $V$-band magnitude and a 
parallax larger than 0.5~mas. Fig.~\ref{fig_parerr} shows the 
distribution of errors $\sigma\pi$ for these data.
From this dataset the following recipe was devised.

The median parallax error (in mas) is calculated from:
\begin{equation}
{\sigma}_{\pi}^{\rm med}  = 0.914 \; \left( 
\frac{V}{10}\right)^{6.1827}  + 0.735 \,\,\,,
\end{equation}
and the minimum  parallax error (in mas) from:
\begin{equation}
{\sigma}_{\pi}^{\rm min}  = 0.516 \; \left( 
\frac{V}{10}\right)^{5.6010} + 0.451 \,\,\,.
\end{equation}

Then a random number is drawn from a Gaussian distribution with a mean 
of 1.0 and a sigma of 0.180. This number is multiplied with
${\sigma}_{\pi}^{\rm med}$. If this value is larger than
${\sigma}_{\pi}^{\rm min}$ then this is accepted as the parallax
error $\sigma\pi$. If not, (a) new random number(s) is (are) drawn 
until this is fulfilled.

Once a star of magnitude $V$ and his error $\sigma\pi$ are simulated,
the individual measurement error $\delta\pi$ is drawn
from the Gaussian of dispersion $\sigma\pi$.

\end{document}